\documentclass[twoside,12pt]{article}
\usepackage{rotating}
\usepackage{epsfig}

\def\Journal#1#2#3#4{{#1} {#2} (#4) #3 }

\def\NPA{{\em Nucl. Phys.} A}

\def\PLB{{\em Phys. Lett.} B}

\def\PRL{\em Phys. Rev. Lett.}

\def\PREP{\em Phys. Rep.}

\def\PRC{{\em Phys. Rev.} C}

\def\ZPA{{\em Z. Phys.} A}
\def\EPJA{{\em Eur. Phys. J.} A}
\def\EPJAX{{\em Eur. Phys. J. A, accepted}}
\def\ANNP{\em Ann. Phys.}

\def\CNPP{{Com. Nucl. Part. Phys.}}

\def\NIMA{{\em Nucl. Instr. Meth.} A}
\def\IEEE{{\em IEEE Trans. on Nucl. Science}}
\def\PPNP{{\em Prog. in Part. and Nucl. Phys.}}
\begin{document}

\renewcommand{\textfraction}{0.00000000001}
\renewcommand{\floatpagefraction}{1.0}

\title{ \vspace{1cm} Photoproduction of mesons from nuclei\\ - In-medium
properties of hadrons}
\author{B.\ Krusche,$^{1}$\\ 
\\
$^1$Department of Physics and Astronomy, University of Basel, Switzerland\\
}
\maketitle
\begin{abstract} Recent experimental results for the in-medium properties of
hadrons obtained with photoproduction of mesons from nuclei are discussed. 
The experiments were done with the TAPS detector at the tagged photon beam of 
the MAMI accelerator in Mainz. Measured were the final states $\pi^o X$, 
$\eta X$, $2\pi^oX$, and $\pi^o\pi^{\pm}X$ for $^{12}$C, $^{40}$Ca, $^{93}$Nb, 
and $^{208}$Pb up to the second resonance region. The results were used for an 
investigation of the in-medium properties of the P$_{33}$(1232), 
the P$_{11}$(1440), the D$_{13}$(1520), and the S$_{11}$(1535) resonances. 
It was found that the cross sections can be spilt into a component which 
originates from the low density surface region of the nuclei and a component 
which scales like the nuclear volume. The energy dependence of the surface 
component is strikingly similar to the deuteron, it shows a clear signal for 
the second resonance peak. The volume component is lacking this peak and 
shows an enhancement at intermediate energies. Furthermore the 
measurement of coherent $\eta$-photoproduction and the final state $p\pi^o$ 
from $^3$He is discussed in the context of the search for $\eta$-mesic nuclei.
\end{abstract}
\section{Introduction}
In-medium properties of hadrons are a hotly debated topic since they are 
closely connected to the properties of low energy non-perturbative QCD.
QCD at high energies or short scales ($r < 0.1$ fm) is a perturbative theory
with point-like quarks and gluons. However, at larger distances 
the perturbative picture breaks down. In the intermediate range (0.1 fm $< r <$
1 fm) the physics is governed by the excitation of nucleon resonances. This is
the regime where the full complexity of the structure of the nucleon as 
a many body system with valence quarks, sea quarks and gluons contributes.
Apart from lattice gauge calculations, so far only models with effective 
degrees of freedom such as constituent quarks and flux tubes are applicable  
to this problem. At even larger distances beyond 1 fm, QCD becomes the theory
of nucleons and mesons (pions) and can be treated in the framework of chiral
perturbation theory. Chiral symmetry is at the very heart of low energy QCD.
In the limit of vanishing current quark masses the chiral Lagrangian is
invariant under chiral rotations, right- and left-handedness of quarks is
conserved and right- and left-handed fields can be treated independently.
The explicit breaking of chiral symmetry due to the finite u,d current quark
masses (5-15 MeV) is small. However, it is well known, that chiral symmetry is
spontaneously broken since the ground state, the QCD vacuum, has only part
of the symmetry of the Lagrangian. This is connected with a non-zero
expectation value of scalar $q\bar{q}$ pairs in the vacuum, the so-called
chiral condensate. A consequence of the chiral symmetry breaking in the
hadron spectra is the non-degeneracy of parity doublets. The $J^{\pi}=0^-$
pion (the Goldstone boson of chiral symmetry) is much lighter than its chiral
partner the $J^{\pi}=0^+$ $\sigma$ meson. Similarly, the lowest lying 
$J^{\pi}=1^-$ meson, the $\rho$ has a smaller mass than the $J^{\pi}=1^+$
$a_1$ and also the first $J^{\pi}=1/2^-$ excited state in the baryon spectrum, 
the S$_{11}$(1535), lies much above the $J^{\pi}=1/2^+$ nucleon ground state.

\begin{figure}[h]
\hspace*{9cm}\begin{minipage}{9. cm}
\epsfysize=6.3cm \epsffile{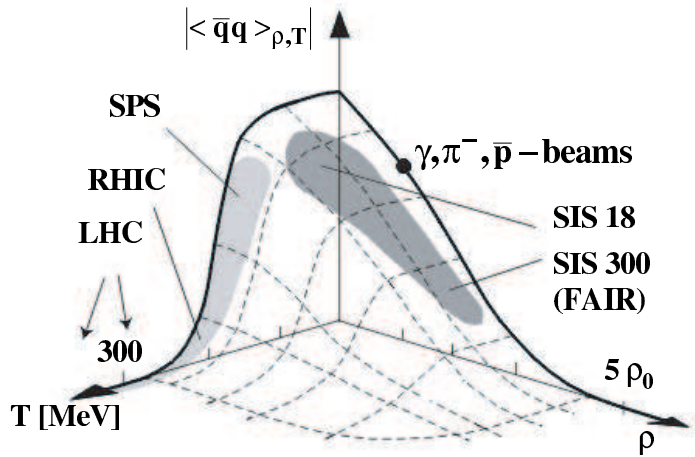}
\caption{Chiral condensate as function of temperature $T$ and nuclear density
$\rho$ ($\rho_o$ normal nuclear matter density)}
\label{fig:1}
\end{minipage}
\end{figure}

\vspace*{0.5cm}
\noindent
\begin{minipage}{8cm}
Model calculations indicate a significant temperature and density dependence 
of the chiral condensate (see e.g. \cite{Lutz_92}). This behavior is 
illustrated in fig. \ref{fig:1}. The melting of the chiral condensate
is connected with a predicted partial restoration of chiral symmetry at high
temperatures and/or large densities. The different regimes are in particular
accessible in heavy ion reactions, but the effect is already significant
at zero temperature and normal nuclear matter density, i.e. conditions which 
can be probed with photon and pion beams. One consequence of the partial chiral
symmetry restoration is a density dependence of hadron masses. An early
prediction for this effect is the so-called Brown-Rho scaling \cite{BR_91}: 
\begin{equation}
m^{\star}_{\sigma ,\rho ,\omega}/m_{\sigma ,\rho ,\omega} \approx
m^{\star}_N/m_N \approx f^{\star}_{\pi}/f_{\pi},
\end{equation}  
\end{minipage}

\vspace*{-9.5cm}

\vspace*{0.5cm}
\noindent{where} $m^{\star}$ are the in-medium masses and $f_{\pi}$ is the 
pion decay constant. Simple approximations of the mass dependence, for example 
in the framework of the linear sigma model, parameterize it linearly in the 
nuclear density \cite{Hatsuda_99}:
\begin{equation}
m^{\star}_{\sigma ,\rho}\approx m_{\sigma ,\rho}\left(1-\alpha_{\sigma ,\rho}
\frac{\rho}{\rho_o}\right)
\end{equation}
with $\alpha$ in the range 0.2 - 0.3 . Evidence for such effects has been
searched for in many experiments. An example is the search for the predicted
shift and broadening of the $\rho$-meson in the di-lepton spectra of heavy
ion reactions with CERES at CERN \cite{Agakichiev_95,Adamova_03} and in the 
near future with the HADES detector at GSI. Heavy ion induced reactions profit
from the relatively large densities reached in the collision phase, but suffer
from the complicated interpretation of the rapidly varying, highly
non-equilibrium reaction conditions. More recently, also pion and photon 
induced reactions on nuclei, which test the hadron properties at normal
nuclear matter density, have moved into the focus. A much discussed effect in
this field is the in-medium modification of the $\sigma$ meson, respectively 
the modification of the $\pi\pi$ interaction in the scalar - isoscalar 
channel, reported from the CHAOS \cite{Bonutti_96} and Crystal Ball 
collaborations {\cite{ Starostin_00} for pion induced double pion production.   

In-medium modifications of mesons will of course also influence the in-medium
properties of nucleon resonances due to the coupling between resonances and 
mesons. Recently, Post, Leupold and Mosel \cite{Post_04} have calculated in a 
self consistent way the spectral functions of mesons and baryons in nuclear 
matter from these couplings. The most relevant contributions to the 
self-energies are shown in fig. \ref{fig:2}. In the vacuum mesons like the 
$\rho$ can couple only to meson loops (involving e.g. the pion) and nucleon 
resonances couple to nucleon - meson loops. However, in the medium mesons can
couple to resonance - hole states (the best known example is the coupling of
the pion to $\Delta -h$ states in $\Delta$ - hole models). This influences not
only the spectral functions of the mesons, but also the resonances which in 
turn couple to the modified meson loops. It makes necessary an iterative,
self-consistent treatment of the self-energies. The predicted effects are in
particular large for the $\rho$ meson and the D$_{13}$(1520) resonance due to
the strong coupling of the resonance to $N\rho$. The close by S$_{11}$(1535)
resonance is much less effected (see \cite{Post_04}).  

\begin{figure}[t]
\epsfysize=2.3cm \epsffile{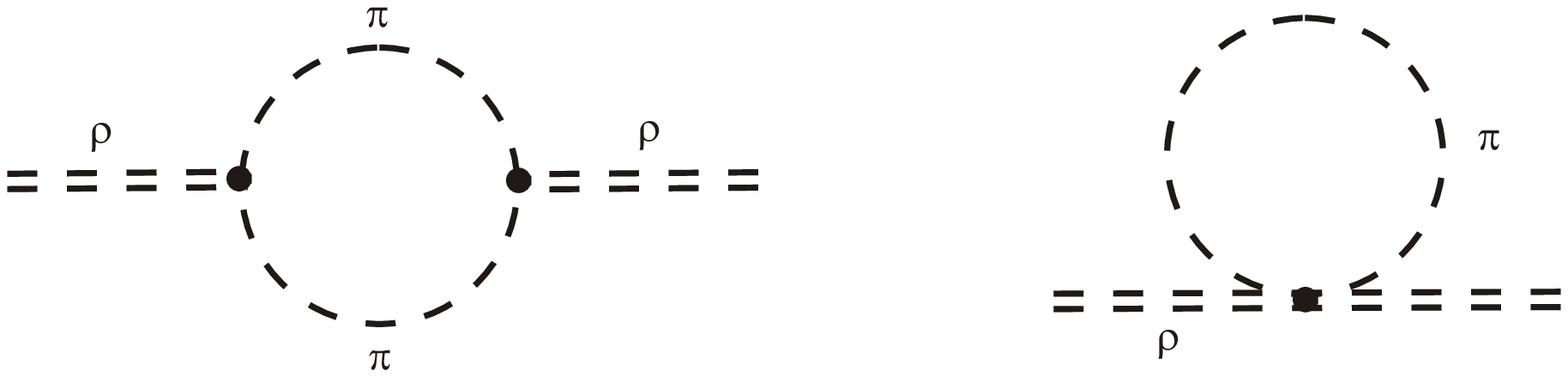}
\hspace*{0.5cm}\epsfysize=2.3cm \epsffile{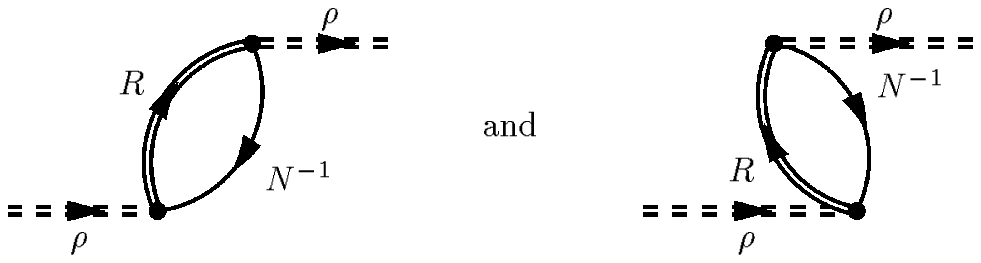}\\
\hspace*{2cm}\epsfysize=2.5cm \epsffile{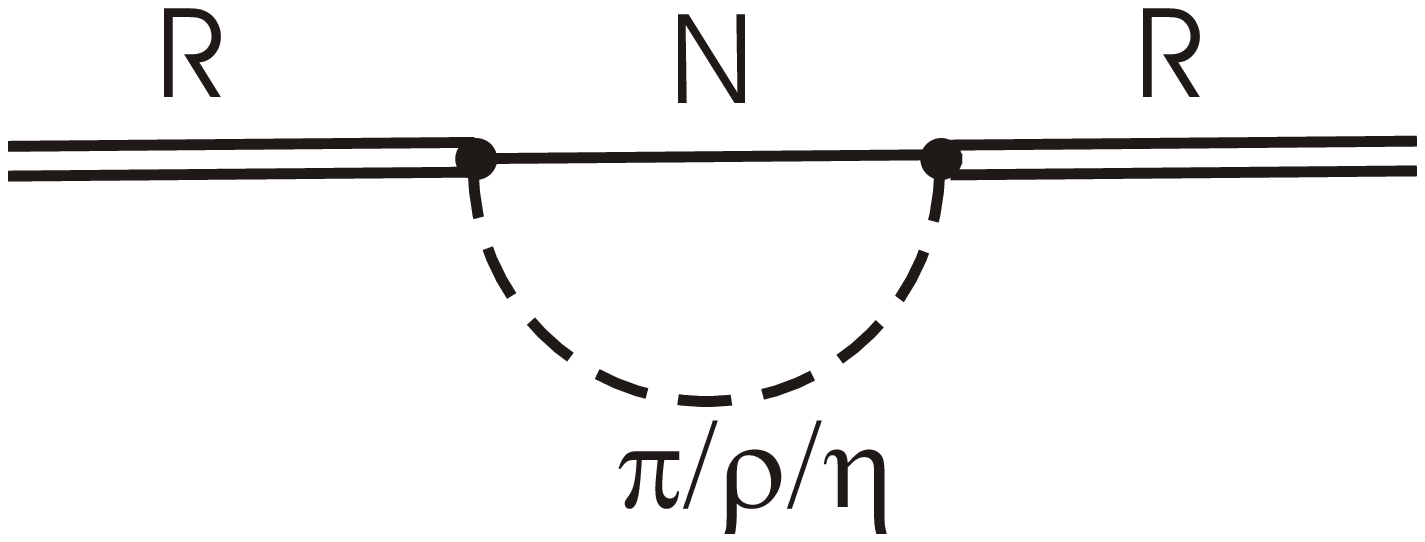}
\hspace*{3cm}\epsfysize=2.5cm \epsffile{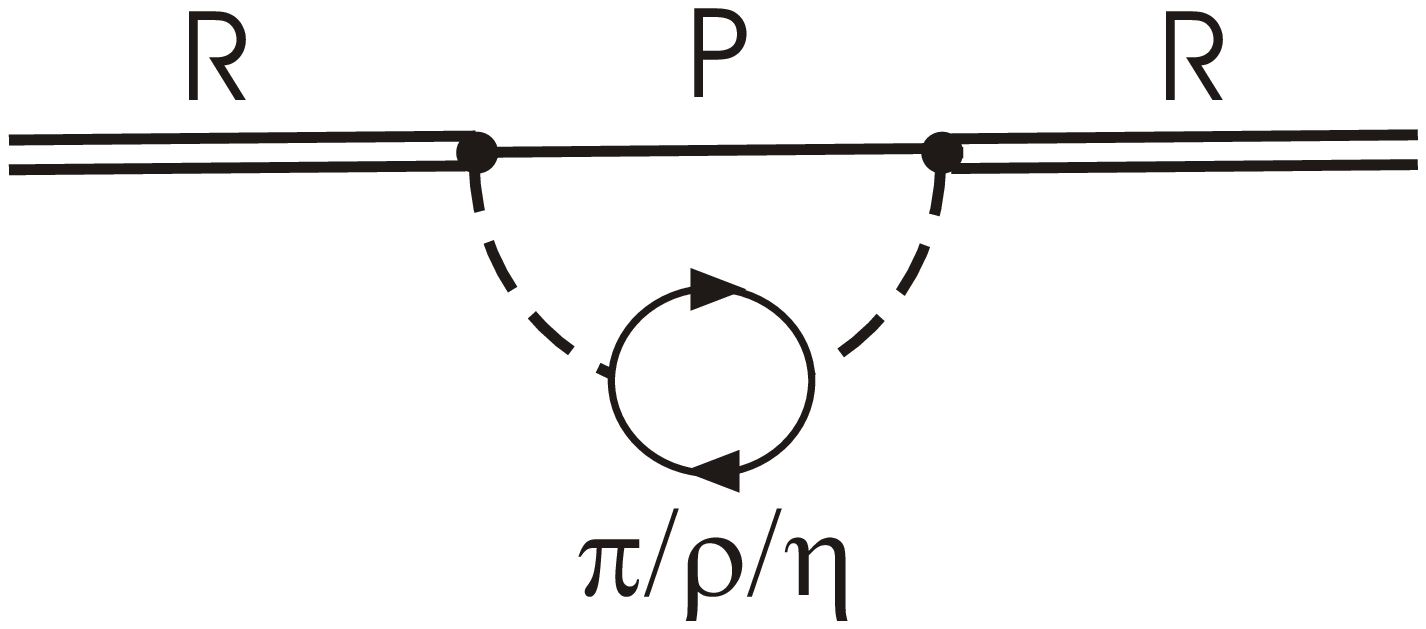}
\caption{Self energies from coupling of mesons and nucleon resonances, left
hand side: vacuum, right hand side: in the nuclear medium \cite{Post_04}}
\label{fig:2}
\end{figure}
                                                                              
During the last few years, the TAPS collaboration has engaged in a program to
study the in-medium properties of mesons and nucleon resonances and the meson -
nucleus interactions with photon induced meson production reactions from 
nuclei. This program covers four different major topics: 
\begin{itemize}
\item{The investigation of resonance contributions to $\eta$, $\pi$, $2\pi$
meson production reactions from nuclei aiming at resonance in-medium properties
like mass and width \cite{Roebig_96,Krusche_04}}
\item{The search for $\eta$ - nucleus bound states (so-called
$\eta$-mesic nuclei), which would be the ideal testing ground for the
investigation of the $\eta$ - nucleus interaction \cite{Pfeiffer_04}}
\item{The investigation of the pion - pion invariant mass distributions
for $2\pi^o$ and $\pi^o\pi^{\pm}$ production from nuclei, 
aiming at the in-medium behavior of the '$\sigma$'-meson \cite{Messchendorp_02}}
\item{The measurement of the resonance shape of the $\omega$ meson in nuclear
matter from its $\pi^o\gamma$ decay}
\end{itemize} 

The first two topics will be discussed in this contribution, the status of
the other two topics is summarized by V. Metag in the proceedings to the same 
conference.

\section{Experiments}
The experiments discussed in this contribution were carried out at the Glasgow 
tagged photon facility installed at the Mainz microton MAMI. The experiments 
used Bremsstrahlung photons produced with the 850 MeV electron beam in a
radiator foil. The standard tagging range covers photon energies between 
50 and 790 MeV, although for many experiments the low energy section of the
tagger is switched off, to allow for higher intensities at high photon energies.
This is possible since the electron beam intensity is limited by the 
fastest counting photomultipliers in the tagger focal plane at intensities
far below the capabilities of the electron machine. The maximum tagged photon 
energies were 820 MeV with a typical focal plane energy resolution of 2 MeV. 
However, since the intrinsic resolution of the magnet is much better 
(roughly 100 keV), the use of 'tagger microscopes' with scintillation counters 
of much smaller width is possible and planned for example for the second
generation experiments searching for $\eta$-mesic nuclei.

\begin{figure}[t]
\hspace*{0.5cm}\begin{minipage}{11.4cm}
\epsfysize=8cm \epsffile{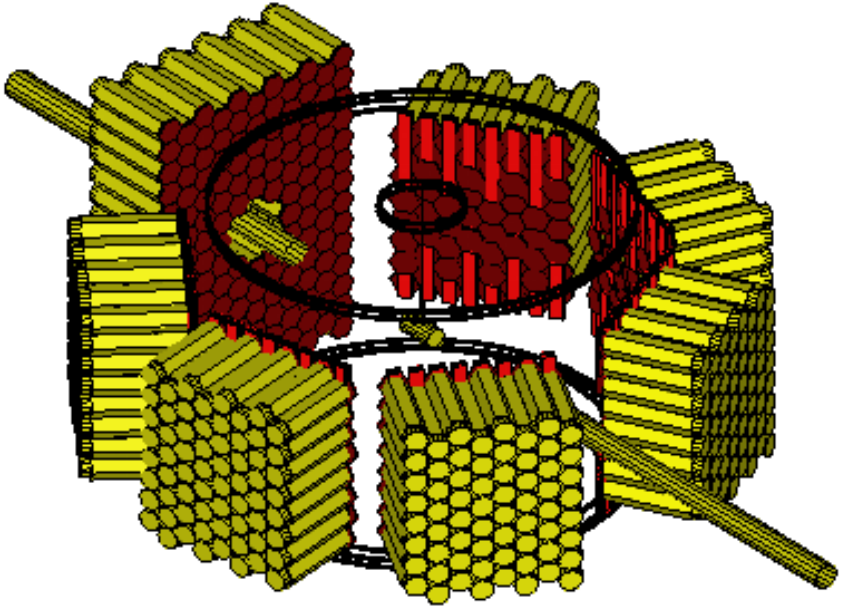}
\end{minipage}
\begin{minipage}{6cm}
\caption{Setup of the TAPS detector at the Mainz MAMI accelerator.
Six block structures with 64 BaF$_{2}$ modules each and one forward wall with
120 crystals were arranged in one plane around the target. The detector modules 
are equipped with individual plastic veto detectors for charged particle 
discrimination. The beam enters from the lower right corner.
}
\label{fig:3}
\end{minipage}
\end{figure}

The meson production experiments were carried out with the electromagnetic
calorimeter TAPS \cite{Novotny_91,Gabler_94}. The setup is shown in fig. 
\ref{fig:3}. It consists of more than 500 hexagonally shaped BaF$_2$ 
scintillators of 25 cm length corresponding to 12 radiation lengths. The 
device is optimized for the detection of photons, but has also particle 
detection capabilities. The separation of photons from massive particles 
makes use of the plastic veto detectors (only charged particles), a 
time-of-flight measurement with typically 500 ps resolution (FWHM) and 
the excellent pulse shape discrimination capabilities of BaF$_2$-scintillators.
The combination of these methods produces extremely clean samples of the 
meson decay photons. The identification of neutral mesons ($\pi^o$ and $\eta$) 
then makes use of a standard invariant mass analysis. Charged mesons and 
nucleons are identified in addition with time-of-flight versus energy 
analyses. Details of the analysis procedures and the identification of 
different reaction channels are summarized in \cite{Krusche_99,Krusche_04}.

\section{Results}
Data have been taken for $^2$H, $^{3,4}$He, $^{12}$C, $^{40}$Ca, $^{93}$Nb and 
$^{208}$Pb targets (the carbon, calcium and lead targets were not isotopic
pure). The data from the deuteron were used as a reference point for the
elementary cross sections from the quasifree nucleon. Compared to the free 
proton this has the advantage that it automatically averages over neutron and 
proton cross section. The measurement with the helium targets were motivated by
the detailed investigation of $\eta$ threshold production from light nuclei
in view of the $\eta$ - nucleus interaction.

\subsection{\it The $\Delta$(1232) resonance}
The excitation of the $\Delta$ resonance and its propagation through the 
nuclear medium have been intensively studied in heavy ion reactions 
\cite{Cassing_90}, in pion, electron, and photon induced reactions
\cite{Lenz_80,Koch_84}. An in-medium broadening at normal nuclear matter density
of roughly 100 MeV has been extracted from pion nucleus scattering experiments
\cite{Hirata_79}. A detailed understanding of the in-medium properties of this
state is necessary for any interpretation of pion photoproduction reactions 
from nuclei. It dominates single pion production in the low energy region up 
to 500 MeV incident photon energies, but it also contributes at higher 
energies via multiple pion production processes and through re-absorption of 
pions. In photon induced reactions on the free proton, single $\pi^o$ 
photoproduction is best suited to study this state. This is demonstrated in 
fig. \ref{fig:4}, with a comparison of the total cross sections for neutral 
and charged pion production from the proton. The background from non-resonant 
contributions is much more pronounced in the charged channel where pion-pole 
and Kroll-Rudermann terms contribute. Such contributions are suppressed for 
neutral pions so that $\gamma p\rightarrow p\pi^o$ is dominated by the 
$\Delta$ already close to threshold.   
\begin{figure}[t]
\epsfysize=7cm \epsffile{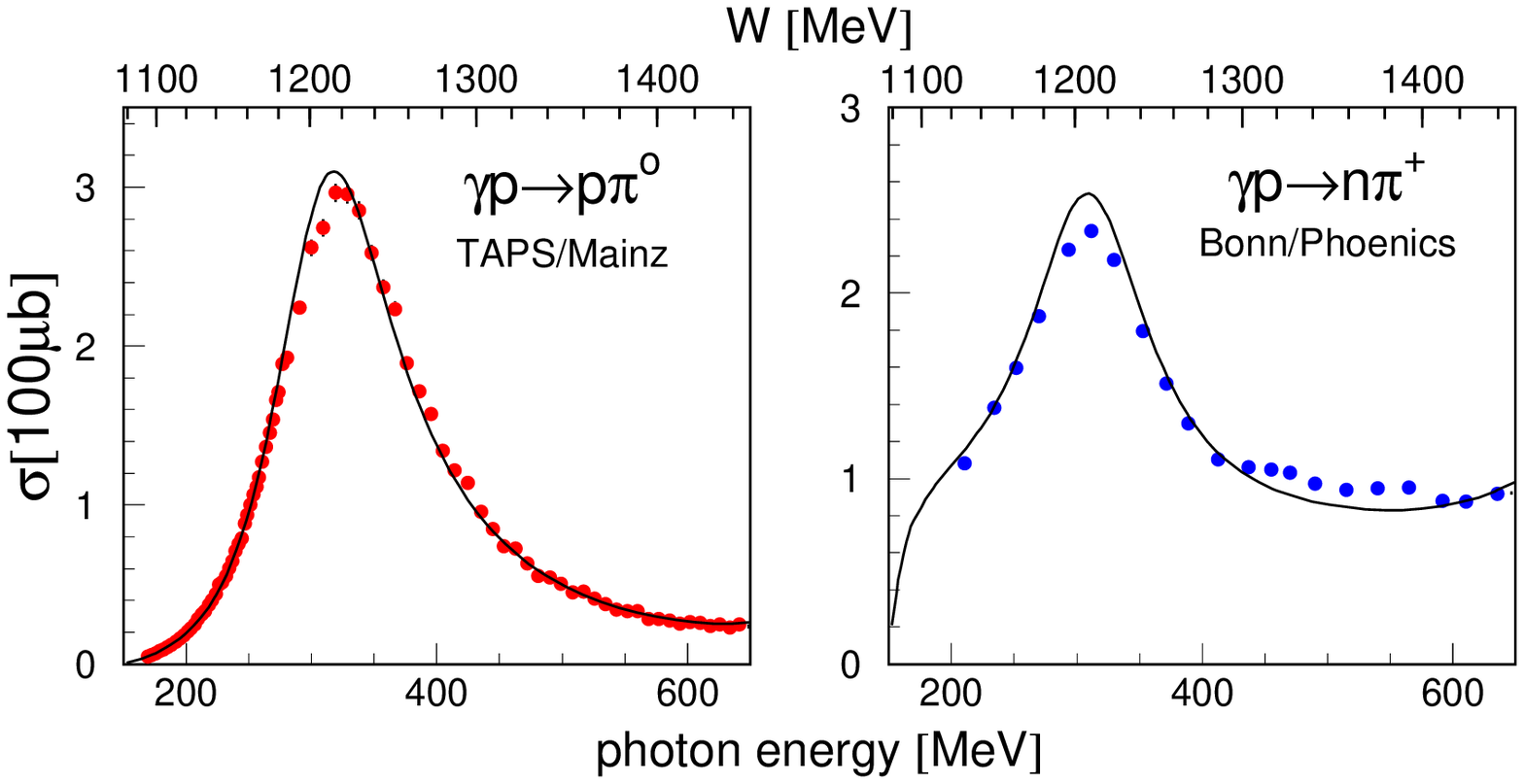}
\hspace*{0.5cm}\epsfysize=6.5cm \epsffile{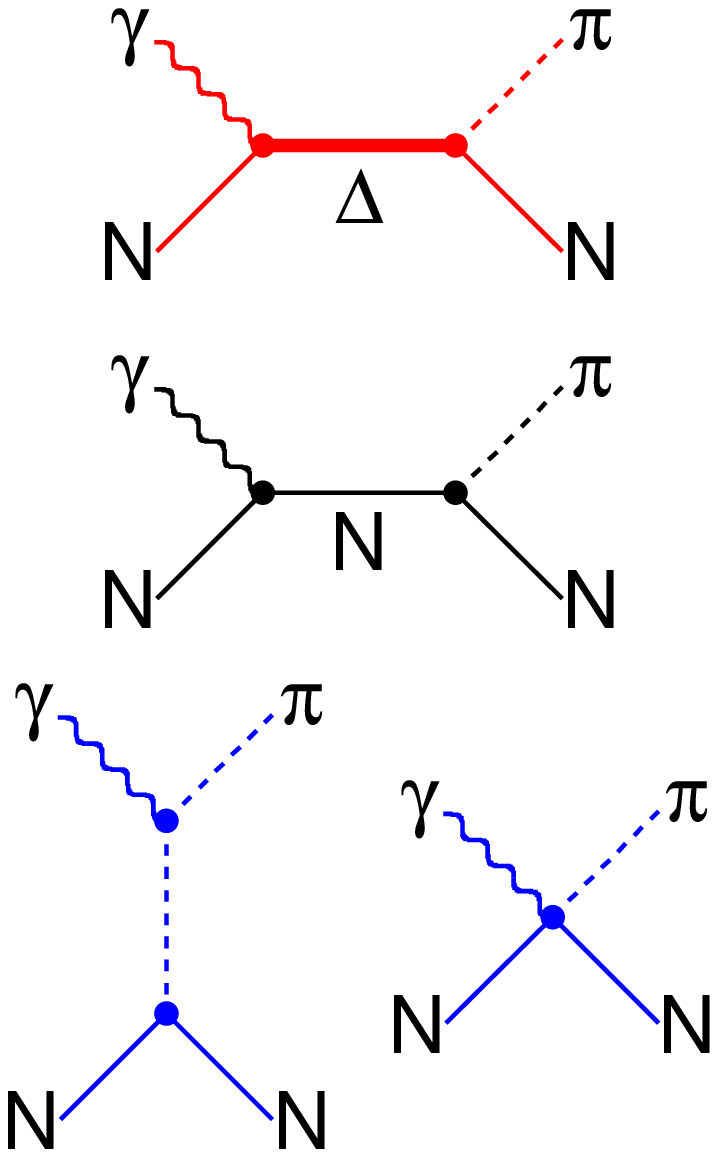}
\caption{Total cross sections for pion photoproduction from the proton in the
$\Delta$-resonance region. Data from \cite{Fuchs_96,Krusche_99} ($p\pi^o$) and
\cite{Buechler_94} ($n\pi^+$). Curves from MAID2000 \cite{MAID_00}.}
\label{fig:4}
\end{figure}
However, for nuclei a further complication
arises. Neutral pions can be produced in two different reaction types
with very different characteristics. In (quasifree) breakup reactions 
in simplest plane wave approximation the pion is produced from a single
nucleon which in the process is knocked out of the nucleus. As long as the
momentum transfer is not too high, this process competes with coherent $\pi^o$
production. In this case the amplitudes from all nucleons add coherently, the
momentum transfer is taken by the entire nucleus, and no nucleons are removed. 
The two reaction mechanisms can be separated via their different kinematics.
The total cross sections for the deuteron and the heavy nuclei are summarized 
for both reaction mechanisms \cite{Krusche_99,Krusche_02,Krusche_04}
in fig. \ref{fig:5}.  
\begin{figure}[ht]
\epsfysize=5.8cm \epsffile{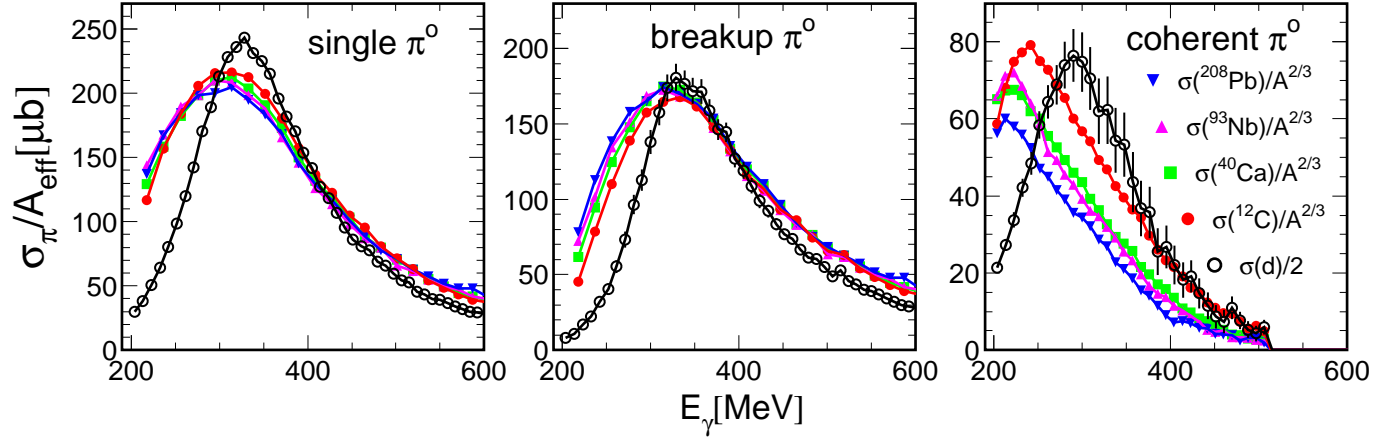}
\caption{Total cross sections for $\pi^o$ production in the $\Delta$ resonance
from the deuteron and from heavy nuclei. Left hand side: coherent part,
middle: breakup part, right hand side: sum of both 
\cite{Krusche_99,Krusche_02,Krusche_04}. The legend is valid for all three 
pictures.}  
\label{fig:5}
\end{figure}
Their behavior is quite different. The coherent reaction can be approximated
for spin $J=0$ nuclei in plane wave by:
\begin{equation}
\frac{d\sigma_A}{d\Omega} \propto 
\frac{d\sigma_N}{d\Omega} A^2 F^2(q) sin^2(\Theta^{\star})
\end{equation}
where $d\sigma_A$ is the nuclear cross section, $d\sigma_N$ the elementary
cross section on the free nucleon, $A$ the atomic mass number, $F^2(q)$ the
nuclear form factor depending on the momentum transfer $q$, and $\Theta^{\star}$
the cm polar angle of the pion (for details see \cite{Krusche_02}). The 
observed shift of the peak cross section to low photon energies for heavy 
nuclei is not related to in-medium effects of the $\Delta$ but
is a simple consequence of the interplay between the $F^2(q)$ and 
$sin^2(\Theta^{\star})$ factors. An extraction of $\Delta$ in-medium properties
from the coherent cross section requires more detailed DWIA calculations (see
below).  

It is tempting to argue, that the breakup process, where the pion is
produced in quasifree kinematics from an individual nucleon, is best suited to
study the $\Delta$ in-medium line-shape. However, quasifree and coherent 
contributions are not independent. They are closely connected via final state 
interaction (FSI), which was discussed in detail for the deuteron in 
\cite{Krusche_99,Siodlaczek_01}. Siodlaczek et al. \cite{Siodlaczek_01} have 
even argued that for the deuteron the effect of FSI in the breakup process is 
just counterbalanced by the coherent process so that the sum of the cross 
sections for the coherent and the breakup part with FSI equals the cross 
section of the pure quasifree process without FSI.  
\begin{figure}[ht]
\begin{minipage}{12cm}
\epsfysize=5.cm \epsffile{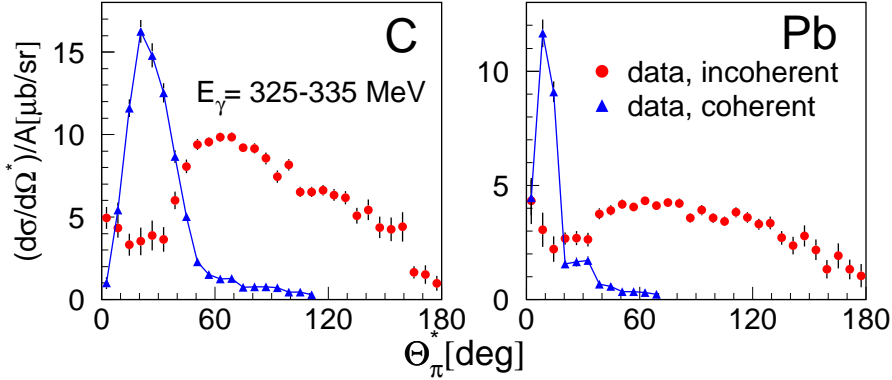}
\end{minipage}
\begin{minipage}{6cm}
\caption{Angular distributions for breakup and coherent $\pi^o$ photoproduction
from carbon and lead \cite{Krusche_04}.}  
\label{fig:6}
\end{minipage}
\end{figure}
As shown in fig. \ref{fig:6} a similar effect is also visible in the angular 
distributions of single pion photoproduction from heavy nuclei
\cite{Krusche_04}. The breakup cross section is depleted at forward angles 
where the coherent cross section peaks. Forward angles of the pion correspond 
to backward angles of the struck nucleon, i.e. to small nucleon momenta
which may lead to Pauli-blocked nucleon final states.
The cross section for inclusive single $\pi^o$ photoproduction, i.e. the sum
of breakup and quasifree parts can thus serve as a first approximation. It
scales almost perfectly with $A^{2/3}$, which of course indicates strong
FSI effects. The average over the heavy nuclei is compared in fig. \ref{fig:7}
to the cross section from the free proton. 
\begin{figure}[ht]
\begin{minipage}{13cm}
\epsfysize=6.3cm \epsffile{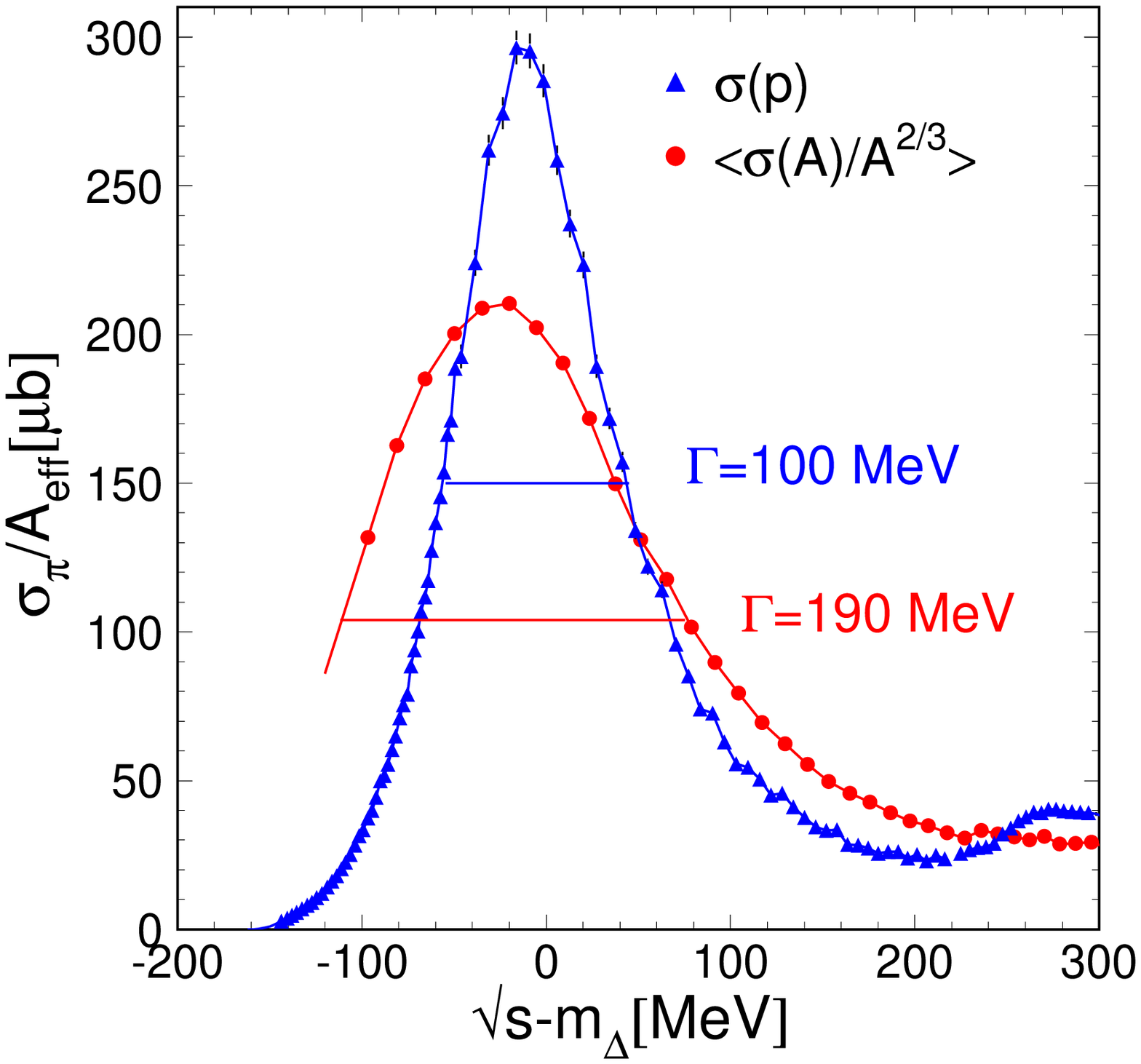}
\hspace*{0.cm}\epsfysize=6.5cm \epsffile{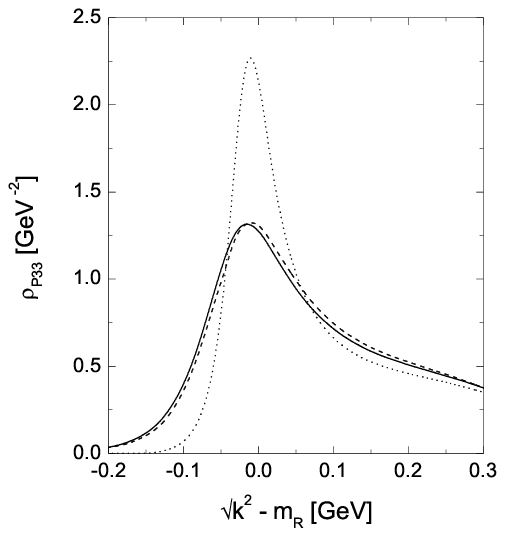}
\end{minipage}
\begin{minipage}{5cm}
\caption{Left hand side: total $\pi^o$ photoproduction from the proton compared
to the average for heavy nuclei. Right hand side: self consistent spectral
functions for the $\Delta$ resonance in vacuum and in the nuclear medium
\cite{Post_04}.}
\label{fig:7}
\end{minipage}
\end{figure}
The $\Delta$-resonance peak for the nuclei is significantly broadened with
respect to the free nucleon from 100 MeV to 190 MeV (note that nuclear Fermi 
motion causes only a much smaller broadening). This is in nice agreement with
the prediction for the in-medium spectral function of the $\Delta$ 
\cite{Post_04}(see fig. \ref{fig:7}, right hand side), which corresponds to
exactly the same broadening.

A more detailed separate analysis of the breakup and coherent components
requires models. The breakup process is mostly treated in the framework of
nuclear cascade models or transport models. The data are compared in fig.
\ref{fig:8} to calculations in the framework of the Boltzmann-Uehling-Uhlenbeck
transport model \cite{Lehr_00,Krusche_04}. The model includes 
an additional in-medium width of the $\Delta$ of roughly 80 MeV at normal
nuclear matter density.
\begin{figure}[ht]
\epsfysize=5.8cm \epsffile{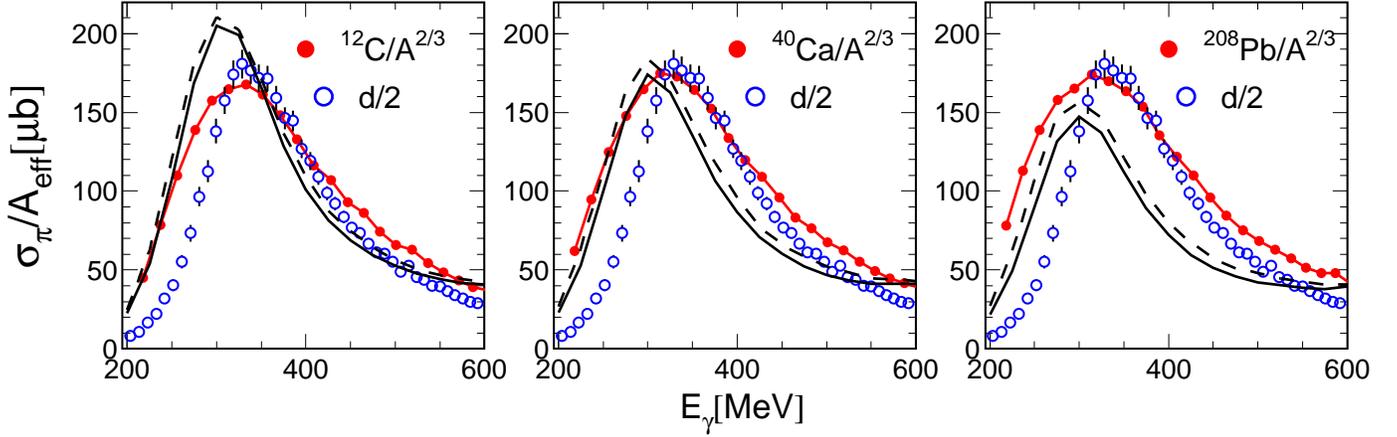}
\caption{Quasifree photoproduction of single $\pi^o$ mesons from the deuteron
and from heavy nuclei compared to BUU-model calculations 
\cite{Lehr_00,Krusche_04} for heavy nuclei. The different curves correspond to
slightly different parameterizations of the $\Delta$ in-medium width.}  
\label{fig:8}
\end{figure}
The calculations reproduce the shift of the rising slope of the $\Delta$ to
lower incident photon energies, but underestimate the falling slope and show a
somewhat different mass number dependence of the peak cross section. Note
however, that for a heavy nucleus like lead FSI reduces the peak cross section
by more than a factor of four, so that the result is extremely sensitive to the
details of FSI in the model. The problems in the high energy tail of the 
$\Delta$ may be partly attributed to two-body absorption processes of the 
photon which are not included in the model. 

The results from coherent $\pi^o$ photoproduction from nuclei have been analyzed
in detail in \cite{Rambo_00,Krusche_02} in the framework of the DWIA
calculations of Drechsel, Tiator, Kamalov and Yang \cite{Drechsel_99a} which
include a phenomenological $\Delta$ self-energy. The result for the total
cross section is shown in fig. \ref{fig:9}.
\begin{figure}[ht]
\begin{minipage}{12.5cm}
\epsfysize=5.3cm \epsffile{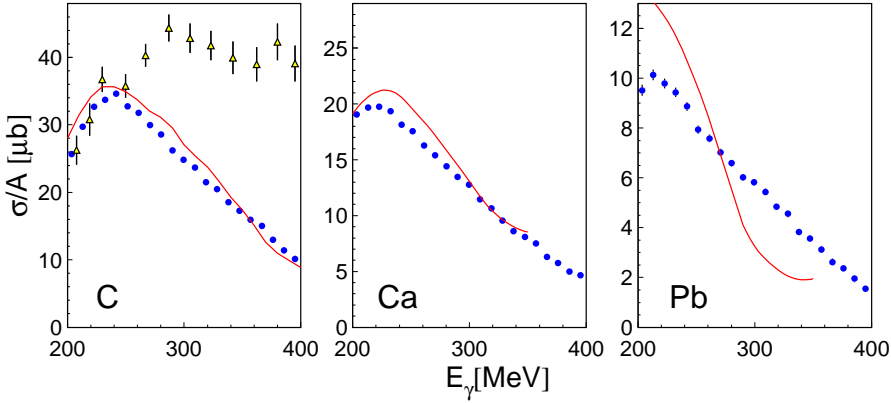}
\end{minipage}
\begin{minipage}{5.5cm}
\caption{Total cross sections for coherent $\pi^o$ photoproduction from nuclei
\cite{Krusche_02}. The curves are calculations from \cite{Drechsel_99a}.
The triangles for $^{12}$C are from a previous measurement with less stringent
suppression of the incoherent components \cite{Arends_83}.}  
\label{fig:9}
\end{minipage}
\end{figure}
The main finding was, that the model with the self-energy fitted to $^4$He
reproduces the data for carbon and calcium almost perfectly so that no
significant mass dependence of the self-energy was found. The self-energy 
itself corresponds to an increase of the width at resonance position
($E_{\gamma}\approx 330$ MeV) of roughly 110 MeV in agreement with the results 
discussed above. The resonance position is slightly upward shifted (by 20 MeV).
This is no contradiction to the excitation functions in fig. \ref{fig:7}. 
The width increase is energy dependent 
(only $\approx$40 MeV at $E_{\gamma}\approx 250$ MeV) so that the net effect in
the excitation functions in fig. \ref{fig:7} is a small downward shift of the
peak position.

\subsection{\it The second resonance region}
Among the clearest experimental observations of in-medium effects is the 
suppression of the second resonance peak in total photoabsorption (TPA) 
experiments \cite{Frommhold_92,Bianchi_93,Bianchi_94}. TPA on the free proton 
shows a peak-like structure at incident photon energies between 600 and 800 MeV, 
which is attributed to the excitation of the P$_{11}$(1440), D$_{13}$(1520), 
and S$_{11}$(1535) resonances. This structure is not visible for nuclei over 
a wide range of mass numbers from lithium to uranium 
(see fig. \ref{fig:10}, left hand side). 
\begin{figure}[ht]
\begin{minipage}{11cm}
\begin{turn}{-90.}
\epsfysize=11cm \epsffile{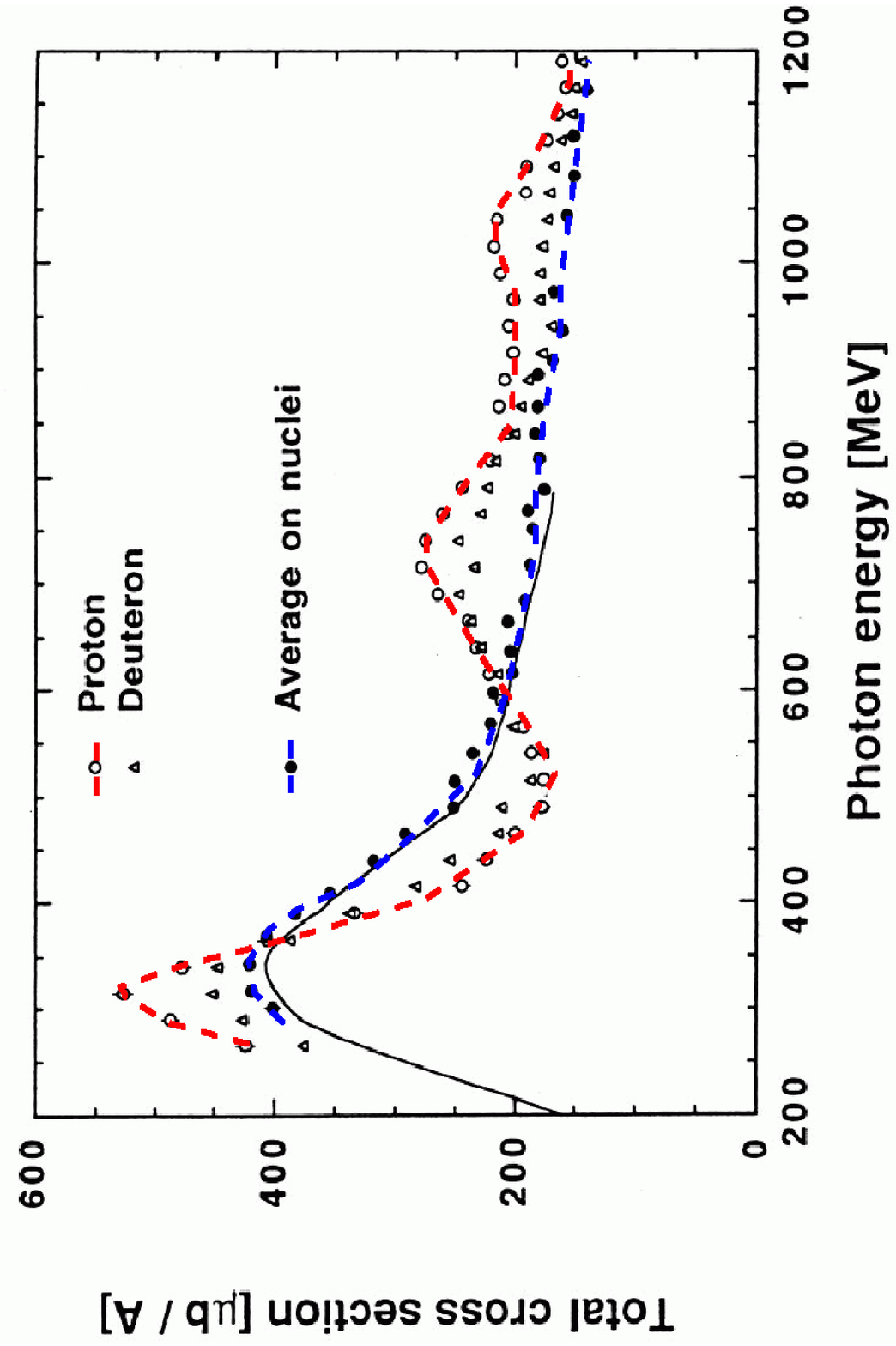}
\end{turn}
\end{minipage}
\begin{minipage}{7.5cm}
\hspace*{0.cm}\epsfysize=7.cm \epsffile{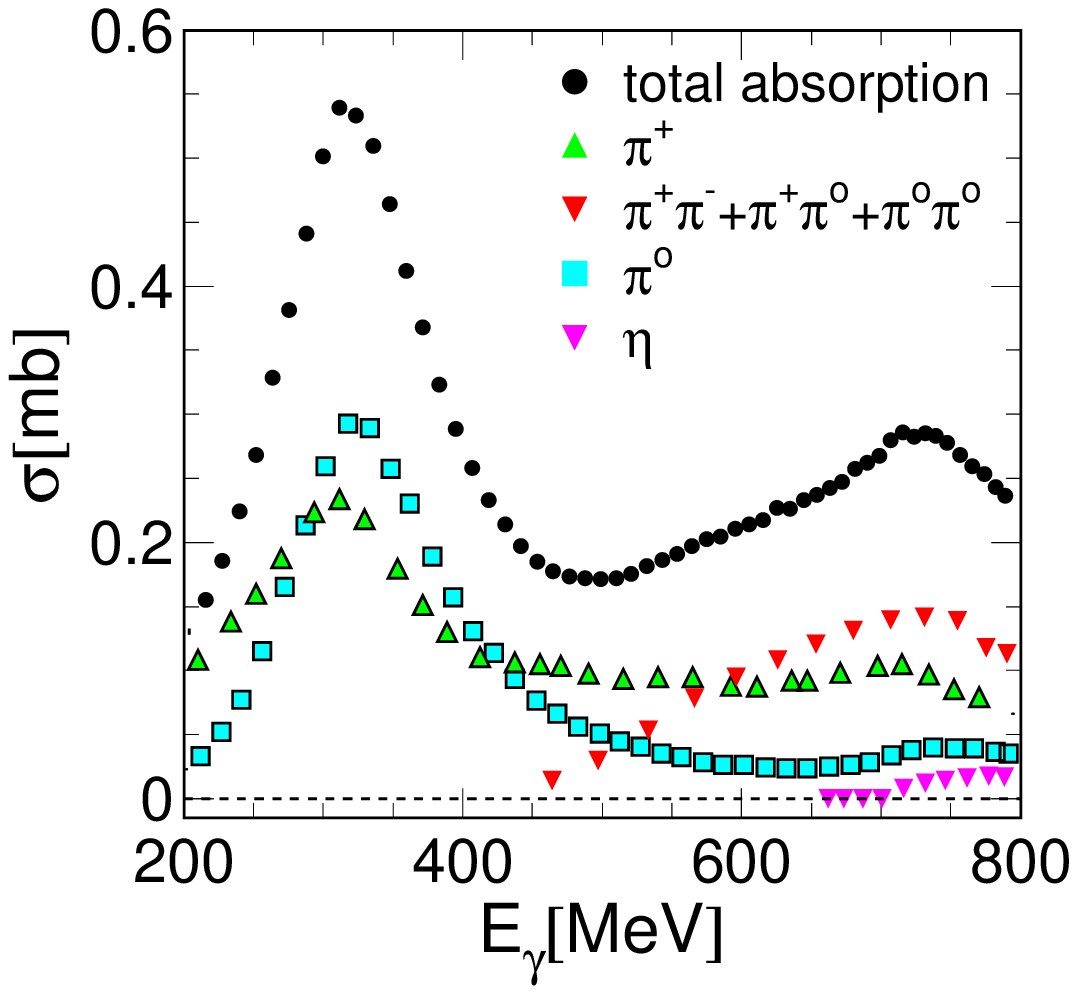}
\end{minipage}
\caption{Left hand side: total photoabsorption from the proton, the deuteron 
and the average for heavy nuclei \cite{Bianchi_94}. Right hand side: Partial 
cross sections for photoproduction from the proton 
\cite{Krusche_99,Buechler_94,Braghieri_95,Krusche_95,MacCormick_96,Haerter_97,
Wolf_00}.}
\label{fig:10}
\end{figure}
A broadening due to nuclear Fermi motion certainly contributes, but cannot 
explain the full effect. Some authors \cite{Kondratyuk_94,Alberico_94} have 
argued for an in-medium width of the relevant 
nucleon resonances, in particular the D$_{13}$(1520), on the order of 300 MeV. 
Such an assumption brings model predictions close to the data but it is not 
clear which effect should be responsible for such a large broadening. 
Post, Leupold and Mosel \cite{Post_04} find in their coupled channel analysis
of in-medium spectral functions of mesons and resonances a relatively 
strong broadening of the D$_{13}$ and a much smaller effect for the S$_{11}$. 
Possible effects resulting from the collisional broadening of the resonances 
have been studied in detail in the framework of transport models of the 
BUU-type (see e.g. \cite{Lehr_00}), but up to now the complete disappearance 
of the resonance structure was not explained.

The resonance bump on the free proton consists of a superposition of reaction 
channels with different energy dependencies (see fig. \ref{fig:10}, right hand
side) which complicates the situation \cite{Krusche_03}. Much of the rise of 
the cross section towards the maximum around 750 MeV is due to the double pion
decay channels, in particular to the  n$\pi^o\pi^+$ and p$\pi^+\pi^-$ final
states. Gomez Tejedor and Oset \cite{Gomez_96} have pointed out that for 
the latter the peaking of the cross section is related to an interference 
between the leading $\Delta$-Kroll-Rudermann term and the sequential decay 
of the D$_{13}$ resonance via D$_{13}\rightarrow\Delta\pi$. Hirata et al. 
\cite{Hirata_98} have argued that the change of this interference 
effect in the nuclear medium is one of the most important reasons for the 
suppression of the bump. 

Inclusive reactions like total photoabsorption alone do not allow
a detailed investigation of such effects. A study of the partial reaction
channels is desirable. The experimental identification of exclusive final 
states is more involved and FSI effects must be accounted 
for. The interpretation of exclusive measurements therefore always needs 
models which account for the trivial in-medium and FSI effects like absorption 
of mesons and propagation of mesons and resonances through nuclear matter. 
On the other hand, as a by-product, the analysis of the FSI effects enables 
the study of meson-nucleus interactions.   

The results for meson photoproduction off the free proton suggest, that
pion and $\eta$ photoproduction are best suited for a comparison of the 
in-medium properties of the D$_{13}$ and S$_{11}$ resonances \cite{Krusche_03}.
This is demonstrated in fig. \ref{fig:11} where the measured cross sections are
compared to the results of the MAID model.
\begin{figure}[h]
\epsfysize=7.7cm \epsffile{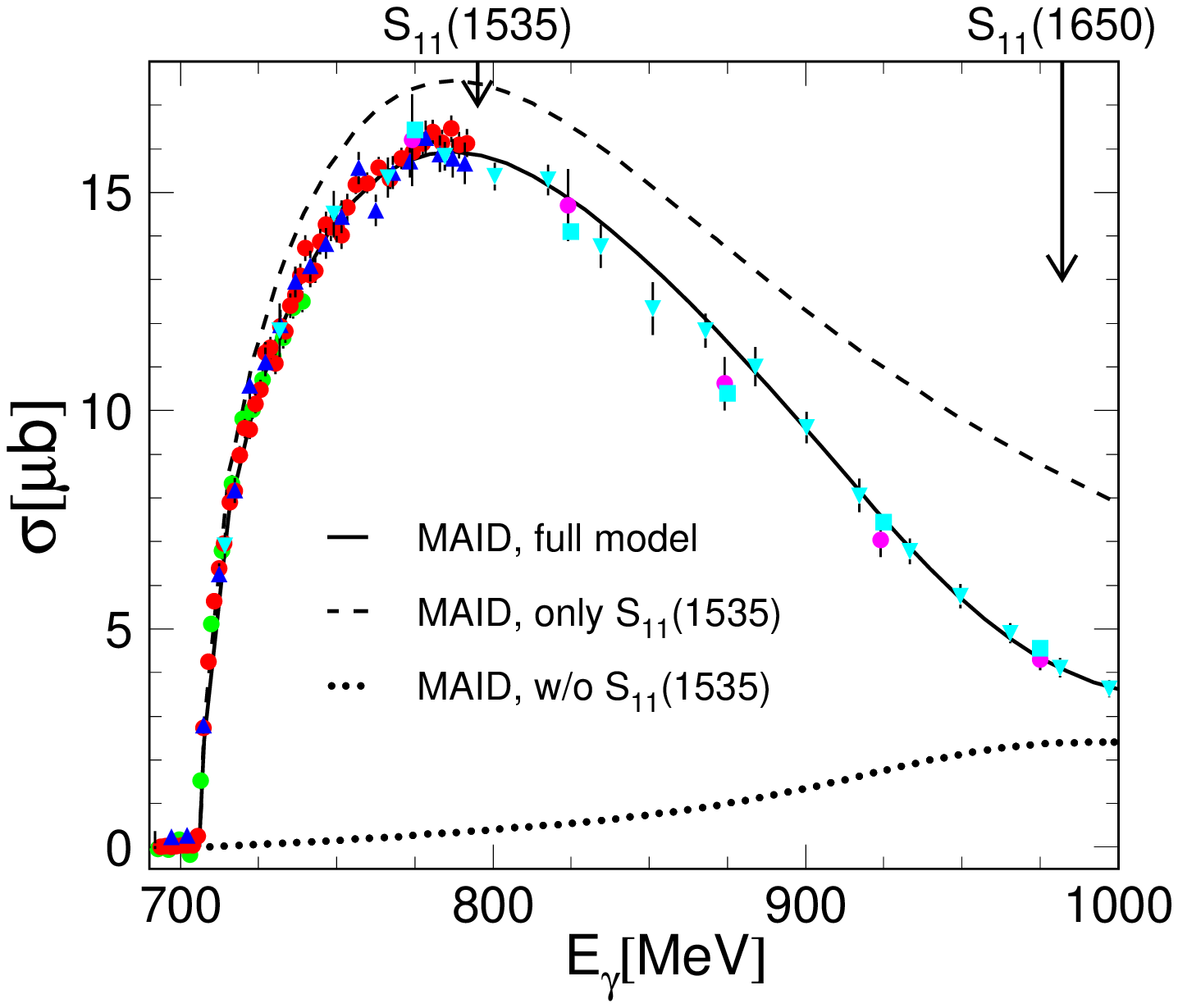}
\hspace*{0.cm}\epsfysize=7.5cm \epsffile{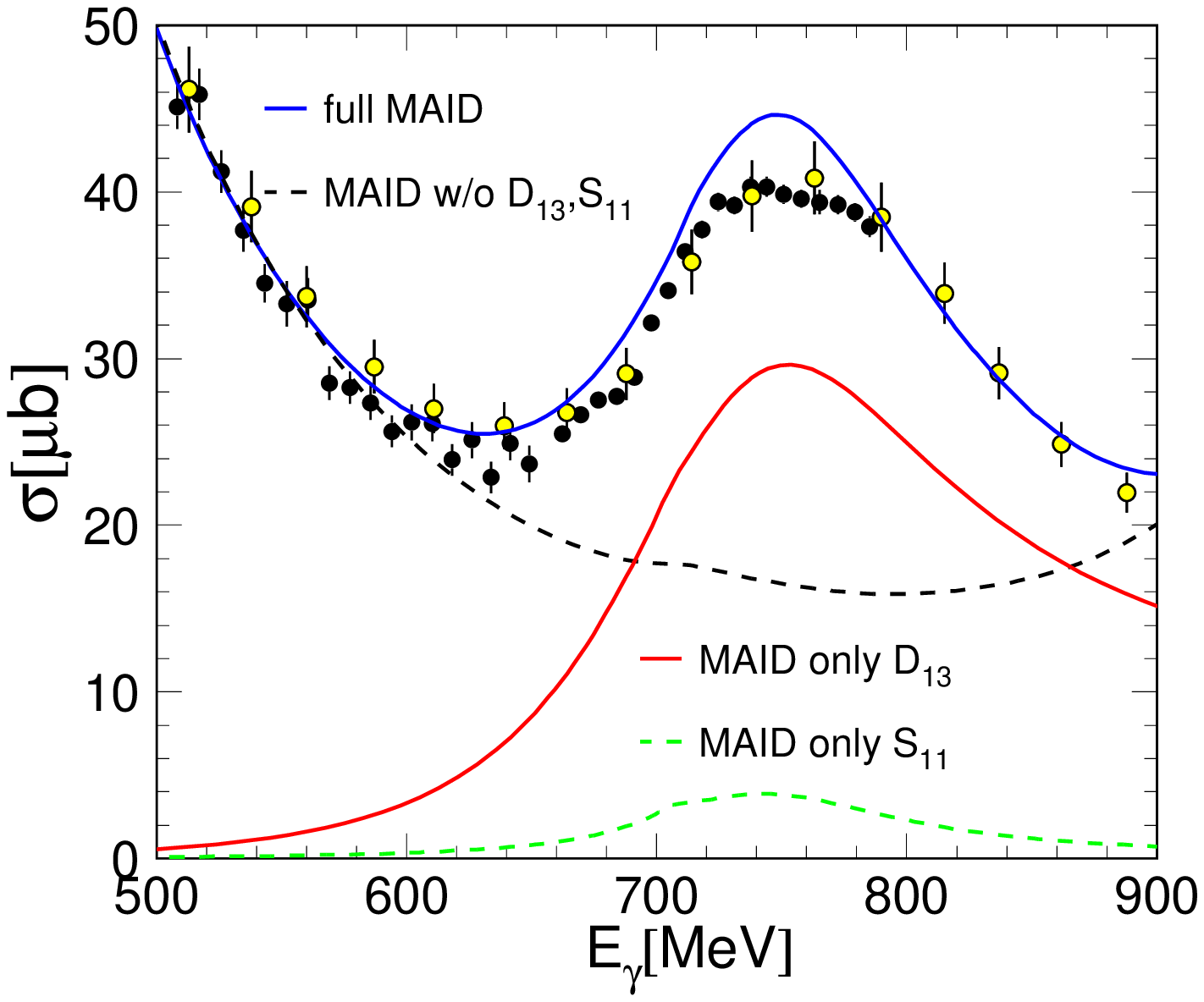}
\caption{Resonance contributions to meson photoproduction off the proton in 
the second resonance region. Left hand side: $\eta$ and the S$_{11}$(1535).
Data from \cite{Krusche_95,Renard_02,Dugger_02,Crede_04}, curves 
from the ETA-MAID model \cite{Chiang_02}. Right hand side: $\pi^o$ and the
D$_{13}$(1520). Data from \cite{Krusche_99,Bartholomy_04} curves from MAID
\cite{MAID_00}.
}
\label{fig:11}
\end{figure}
The total cross section for $\eta$ photoproduction is completely dominated 
in the second resonance region by the S$_{11}$(1535) resonance
\cite{Krusche_97}. The D$_{13}$ has no influence on the total cross section
and only at higher incident photon energies the second S$_{11}$ plays a
significant role. On the other hand, the resonance structure in $\pi^o$ 
photoproduction is strongly dominated by the D$_{13}$(1520) resonance.

Photoproduction of $\eta$ mesons from light to heavy nuclei has been measured 
with TAPS at MAMI \cite{Roebig_96} up to incident photon energies of 800 MeV 
and at KEK in Japan up to 1 GeV \cite{Yorita_00,Yamazaki_00}. A comparison of
the low energy data for the deuteron, $^{12}$C, $^{40}$Ca, $^{93}$Nb and
$^{208}$Pb shows a
perfect scaling with $A^{2/3}$ (the small deviation for the deuteron close to
threshold is due to the smaller Fermi momenta), which indicates strong FSI.
A comparison to calculations with the BUU model \cite{Lehr_00}
or a mean free path Monte Carlo model \cite{Carrasco_93} did not reveal
any in-medium effects for the width or position of the resonance nor any
depletion of its excitation strength \cite{Roebig_96}. A certain drawback of
these data is certainly that they cover the S$_{11}$ resonance only up to its
maximum. The KEK measurements extend somewhat beyond the resonance position
into the downward slope of the excitation functions. The authors claim from a
comparison of the data to quantum Monte Carlo calculations some evidence 
for a broadening of the resonance structure with respect to the elementary
reaction on the free proton. Statistically more precise data over an even much
larger energy range (beyond 2 GeV) have been recently measured with TAPS and 
the Crystal Barrel at the Bonn ELSA accelerator. These data are still under
analysis. However, the preliminary results indicate relatively
large background contributions from $\pi\eta$ final states at incident photon
energies beyond 900 MeV, which have to be suppressed by cuts on the reaction
kinematics. This effect may have contributed to the observed broadening of the
resonance structure in the KEK data, where no kinematical cuts were applied.  
In any way, the claimed in-medium effects for the S$_{11}$ are not large.
A comparison of the data to recent BUU calculations \cite{Lehr_03} 
is shown in fig. \ref{fig:12} right hand side. 
The main result is, that the data are better described when the momentum 
dependence of the in-medium potential for nucleons and the S$_{11}$ is included,
but in agreement with the predicted in-medium spectral function of the S$_{11}$
\cite{Post_04} only small collisional broadening effects are consistent 
with the data. 

The predicted effects for the D$_{13}$ resonance are much larger due to its
strong coupling to the $\rho$ meson \cite{Post_04}. Excitation functions for
single $\pi^o$ photoproduction from the proton, the deuteron and off heavy
nuclei are compared in fig. \ref{fig:13} \cite{Krusche_01}. Surprisingly, the
resonance structure is not significantly broader for the heavy nuclei. 
\begin{figure}[h]
\begin{minipage}{11.8cm}
\epsfysize=8.cm \epsffile{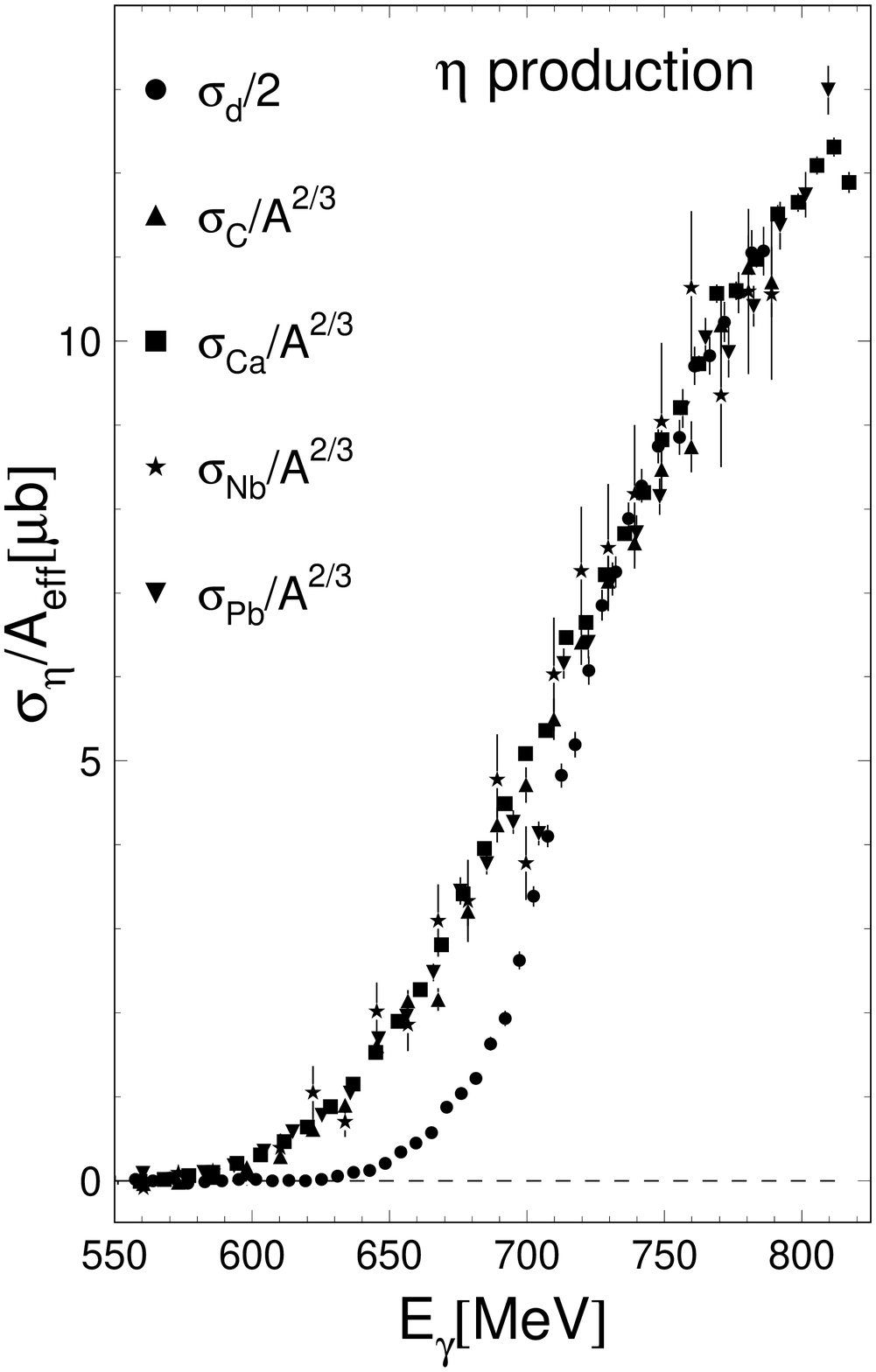}
\epsfysize=8.3cm \epsffile{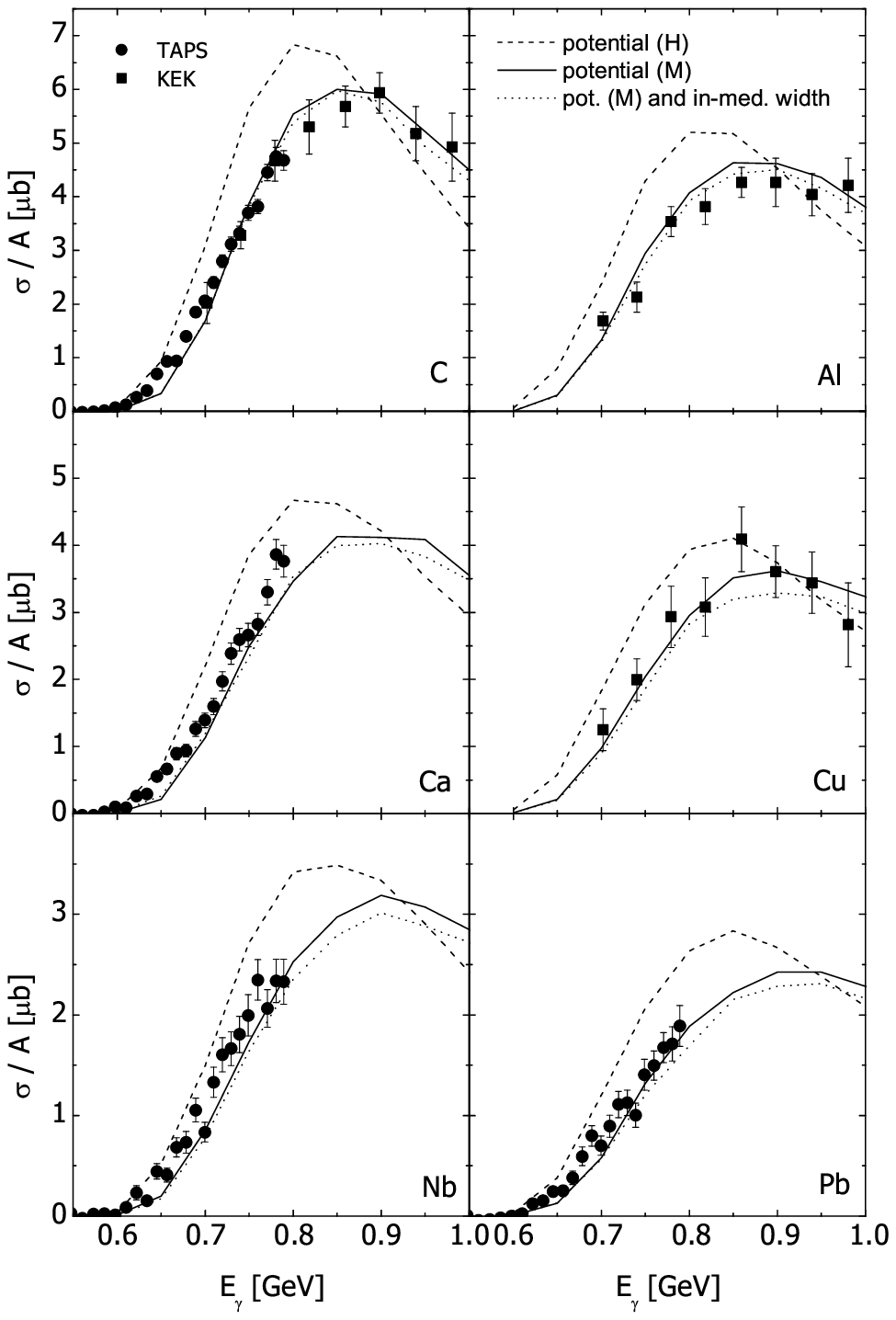}
\end{minipage}
\begin{minipage}{6.5cm}
\caption{Total cross section for $\eta$ photoproduction from nuclei.
Left hand side: scaling of cross sections with mass number. Data from
\cite{Roebig_96,Krusche_04,Krusche_95b,Weiss_03}. Right hand side:
comparison of data \cite{Roebig_96,Yorita_00,Yamazaki_00} to BUU 
calculations \cite{Lehr_03}. Dashed lines: momentum independent potentials,
full lines: momentum dependent potentials, dotted: additional in-medium
broadening of the S$_{11}$
}
\label{fig:12}
\end{minipage}
\end{figure}
\begin{figure}[h]
\begin{minipage}{11.8cm}
\epsfysize=8.cm \epsffile{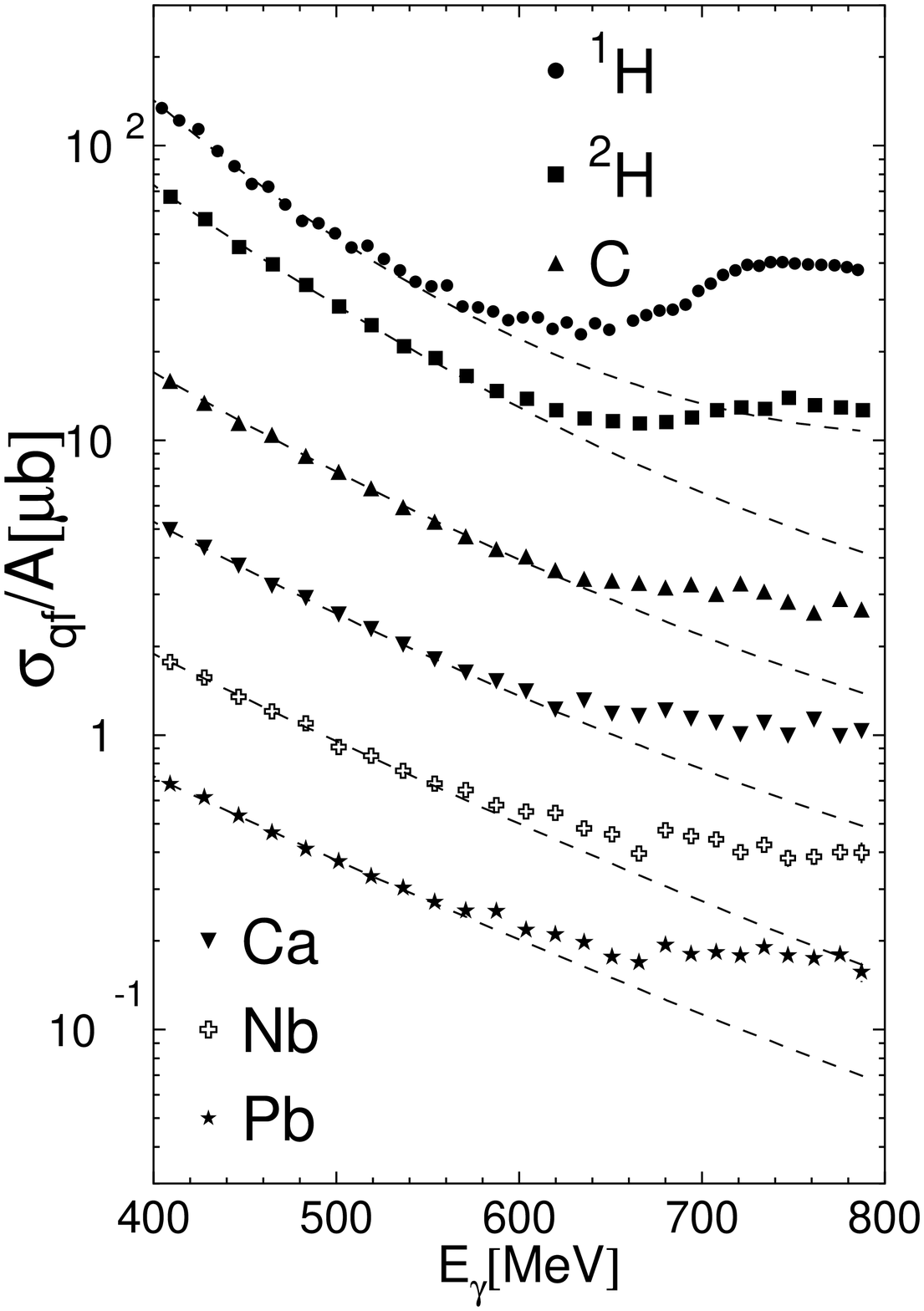}
\epsfysize=8.cm \epsffile{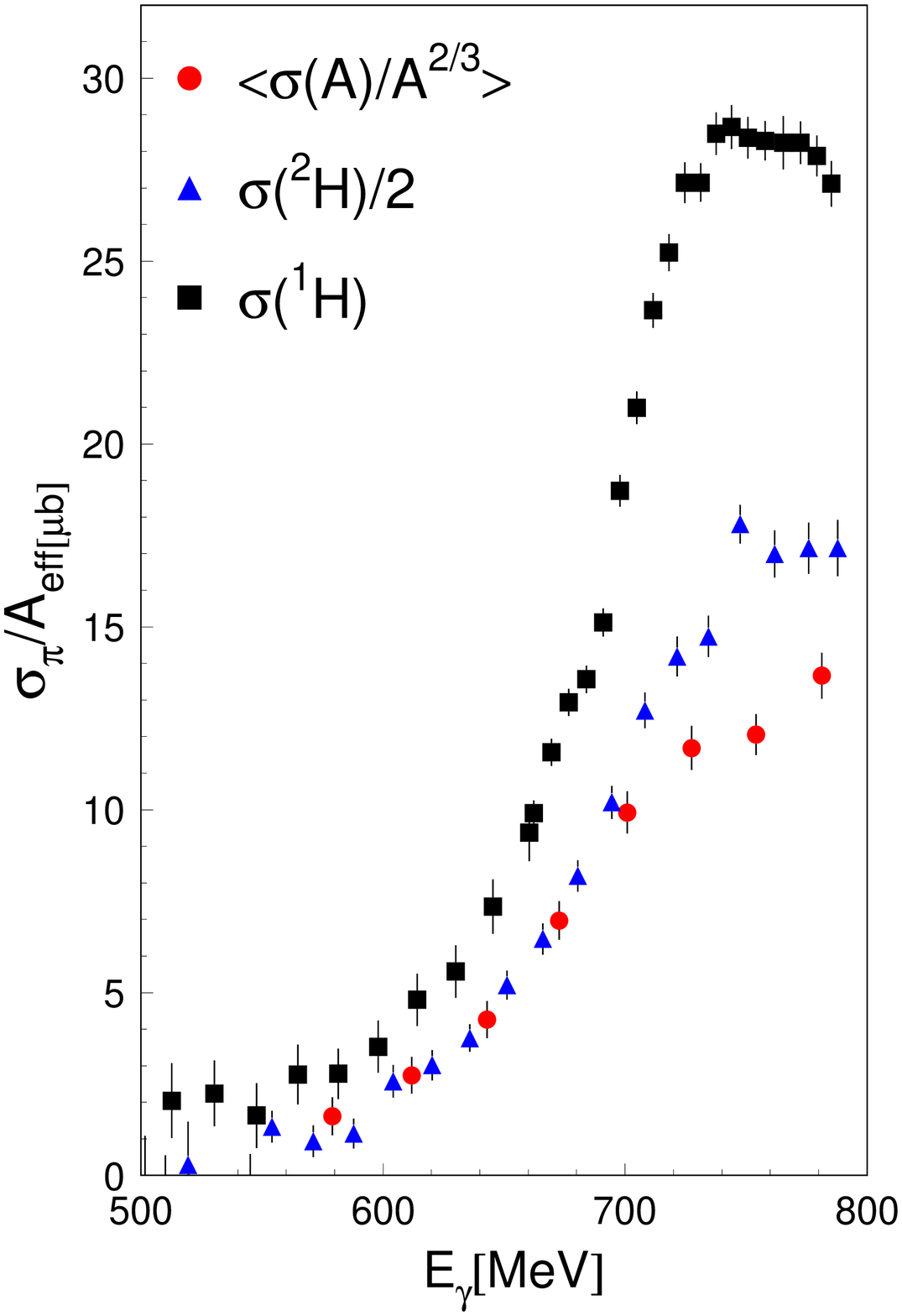}
\end{minipage}
\begin{minipage}{6.5cm}
\caption{Total cross section for single $\pi^o$ photoproduction from nuclei
\cite{Krusche_99,Krusche_01}. 
Left hand side: total cross section with fitted background (fit range 350 - 550
MeV). Scale corresponds to proton data, other data scaled down by 
factors 2,4,8,16, and 32. Right hand side: resonance signal with background
subtracted. For the heavy nuclei only the average is shown. 
}
\label{fig:13}
\end{minipage}
\end{figure}
Also the scaling with the mass number follows the general pattern (see below), 
apart from the fact that a fairly large reduction of the strength occurs for 
the deuteron with respect to the proton. This latter effect is also not yet 
understood \cite{Krusche_99}, predictions for the cross section of 
$n(\gamma,\pi^o)n$ from multipole analyses of pion production are not in
agreement with the observed deuteron cross section.  

A possible explanation for the basically unchanged shape of the D$_{13}$
observed in nuclear pion production could be, that due to FSI, only pions from 
the low density surface region are observed. The absorption properties of
nuclear matter for pions as function of their momentum are summarized in fig.
\ref{fig:14}. 
The left hand side of the figure shows the ratio $R_{C2/3}$ of the cross 
sections for the heavier nuclei and carbon under the assumption of surface 
scaling:
\begin{equation}
R_{C2/3}\equiv\frac{[d\sigma /dp_{\pi}(A)]/A^{2/3}}
{[d\sigma /dp_{\pi}(^{12}C)]/12^{2/3}}\;\;.
\end{equation}
The right hand side shows the scaling exponent $\alpha$ for a power law
scaling $\propto A^{\alpha}$. 
\begin{figure}[h]
\begin{minipage}{13cm}
\epsfysize=6.cm \epsffile{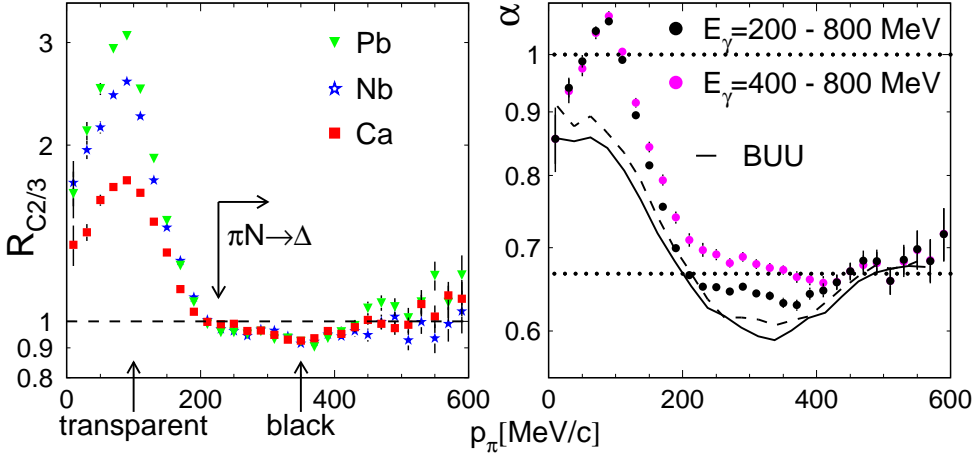}
\end{minipage}
\begin{minipage}{5.4cm}
\caption{Scaling of $\pi^o$ cross sections as function of pion momentum.
Left hand side: ratio  $R_{C2/3}$. Right hand side: scaling coefficient
$\alpha$ determined from $d\sigma /dp_{\pi}\propto A^{\alpha}$ 
\cite{Krusche_04}. Curves from BUU calculations with slightly different
treatment of the $\Delta$ \cite{Lehr_00,Krusche_04}.
}
\label{fig:14}
\end{minipage}
\end{figure}
\begin{figure}[h]
\begin{minipage}{10.3cm}
\epsfysize=9.cm \epsffile{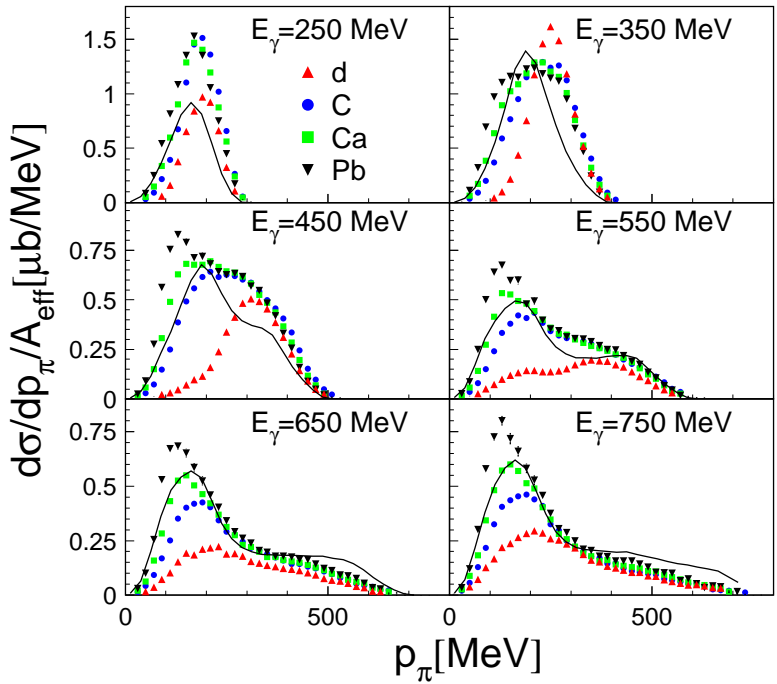}
\caption{Momentum distributions for inclusive $\pi^o$ photoproduction from
nuclei (scaled by $A_{eff}=A^{2/3}$ for $A>2$ and
by $A_{eff}=2$ for the deuteron) \cite{Krusche_04}. Solid lines: BUU
calculations for lead \cite{Lehr_00}.
}
\label{fig:15}
\end{minipage}
\end{figure}

\vspace*{-12cm}
\hspace*{10cm}\begin{minipage}{8cm}
Both pictures demonstrate the expected behavior. Pions with momenta 
large enough to excite the $\Delta$ resonance ($p_{\pi} > 227$ MeV), are 
strongly absorbed ($\alpha\approx$ 2/3). The absorption probability decreases
fast for smaller momenta and the nuclei are almost transparent for pions with
momenta around 100 MeV/c. The influence of FSI and re-scattering can be 
traced in detail in the pion momentum distributions for heavy nuclei, 
which are summarized and compared to the deuteron cross sections and to BUU 
calculations
\cite{Krusche_04,Lehr_00} in fig. \ref{fig:15}. The distributions for the
deuteron approximate the FSI-free case. They show one peak for
single pion production, shifting to higher momenta with increasing incident 
photon energies, and at incident photon energies above 500 MeV a second 
structure at smaller momenta, corresponding to double pion production. 
The qualitative behavior of the spectra for heavier nuclei, reflecting the
re-scattering of pions and the momentum dependence of their mean-free path,
is reproduced by the BUU calculations.  
\end{minipage}

\begin{figure}[h]
\begin{minipage}{13cm}
\epsfysize=4.5cm \epsffile{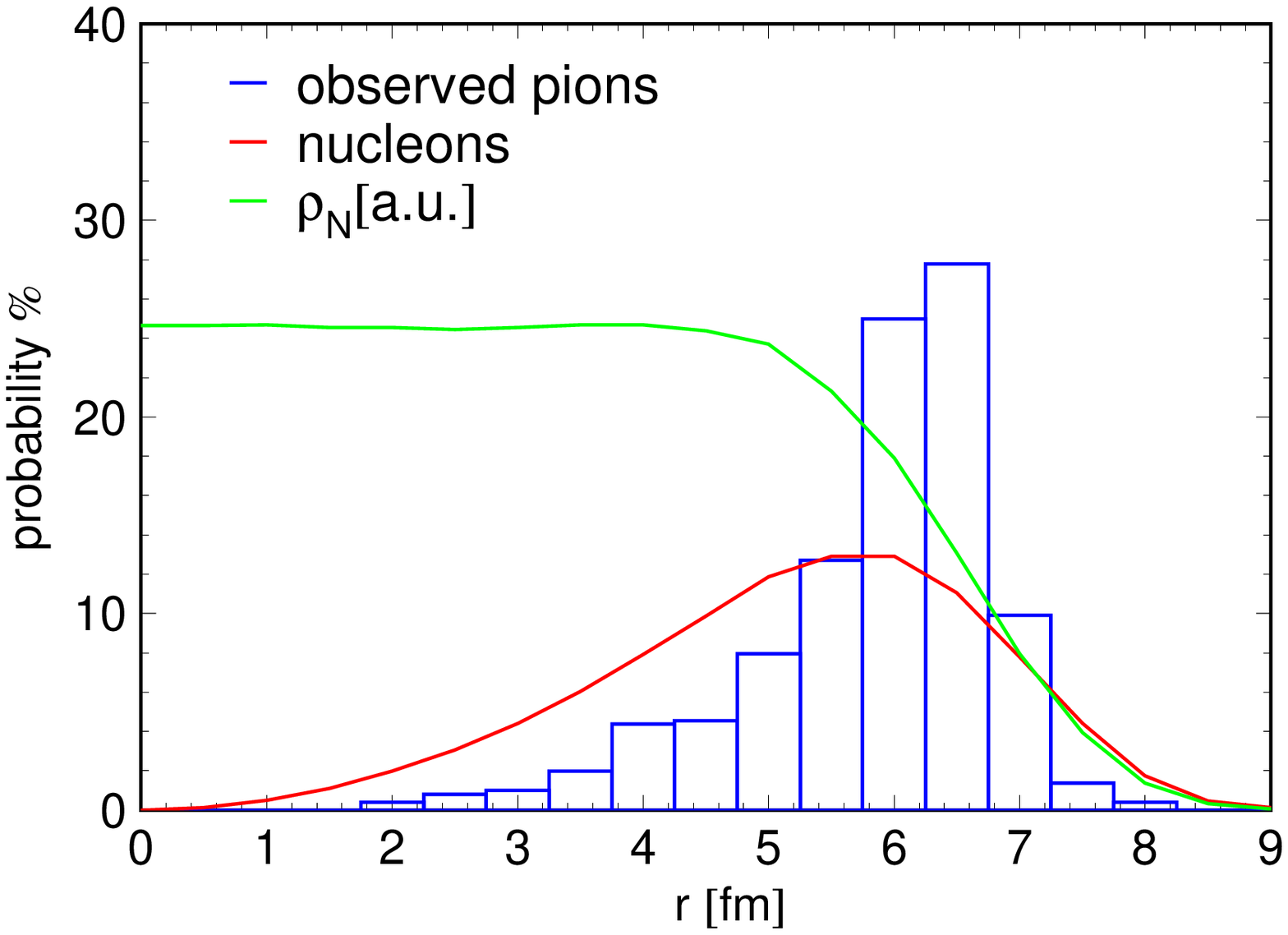}
\epsfysize=4.5cm \epsffile{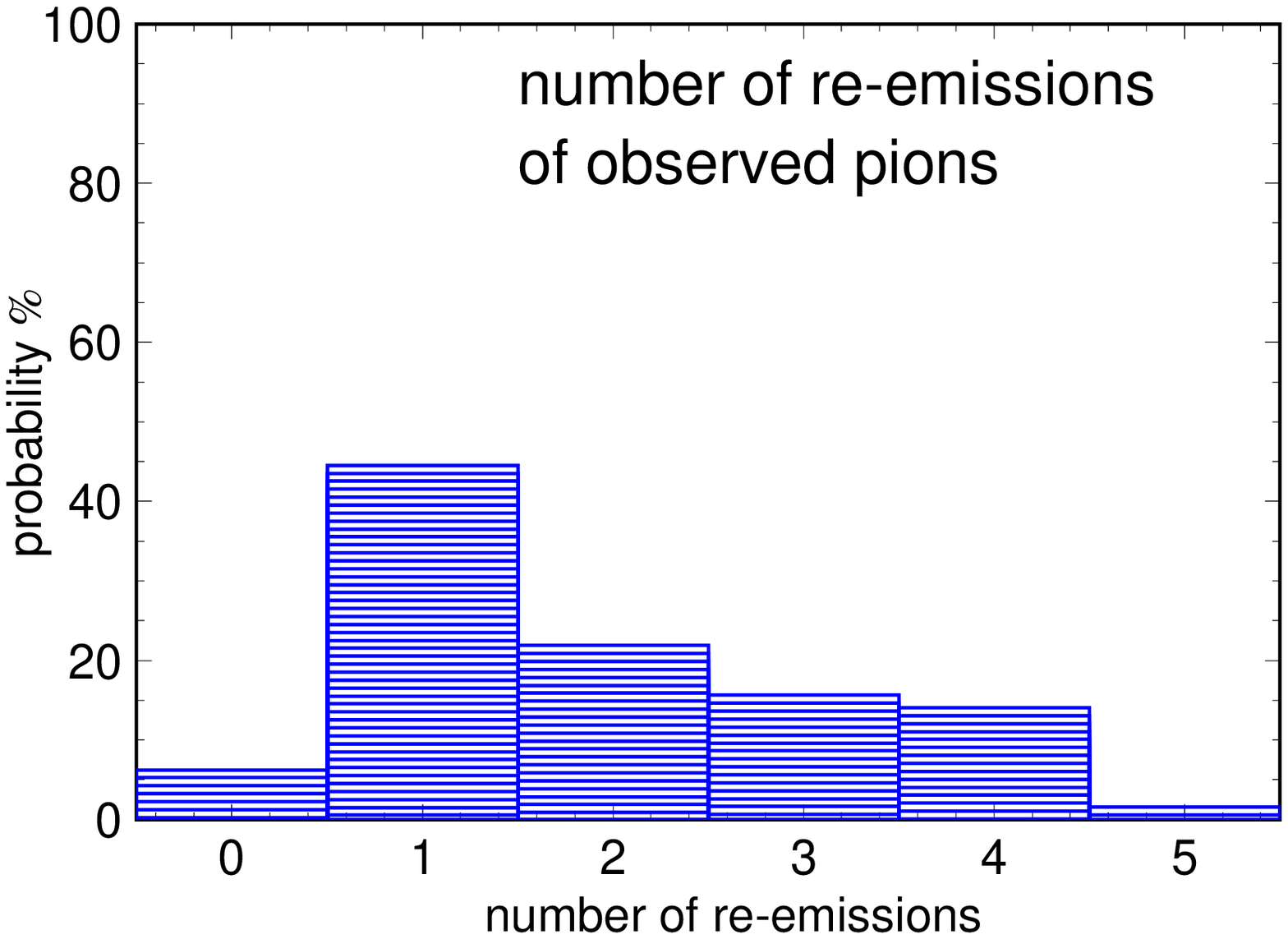}\\
\epsfysize=4.5cm \epsffile{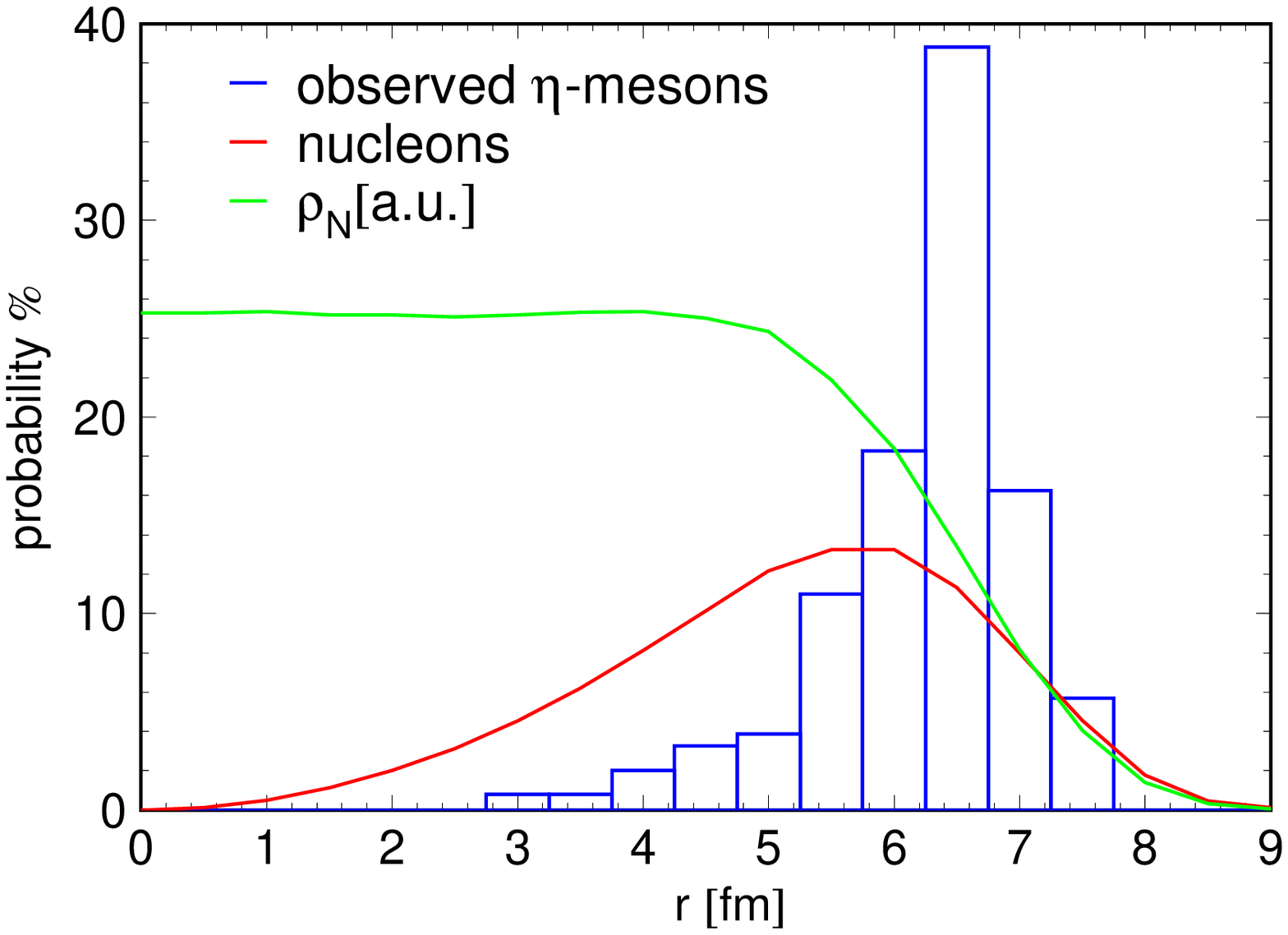}
\epsfysize=4.5cm \epsffile{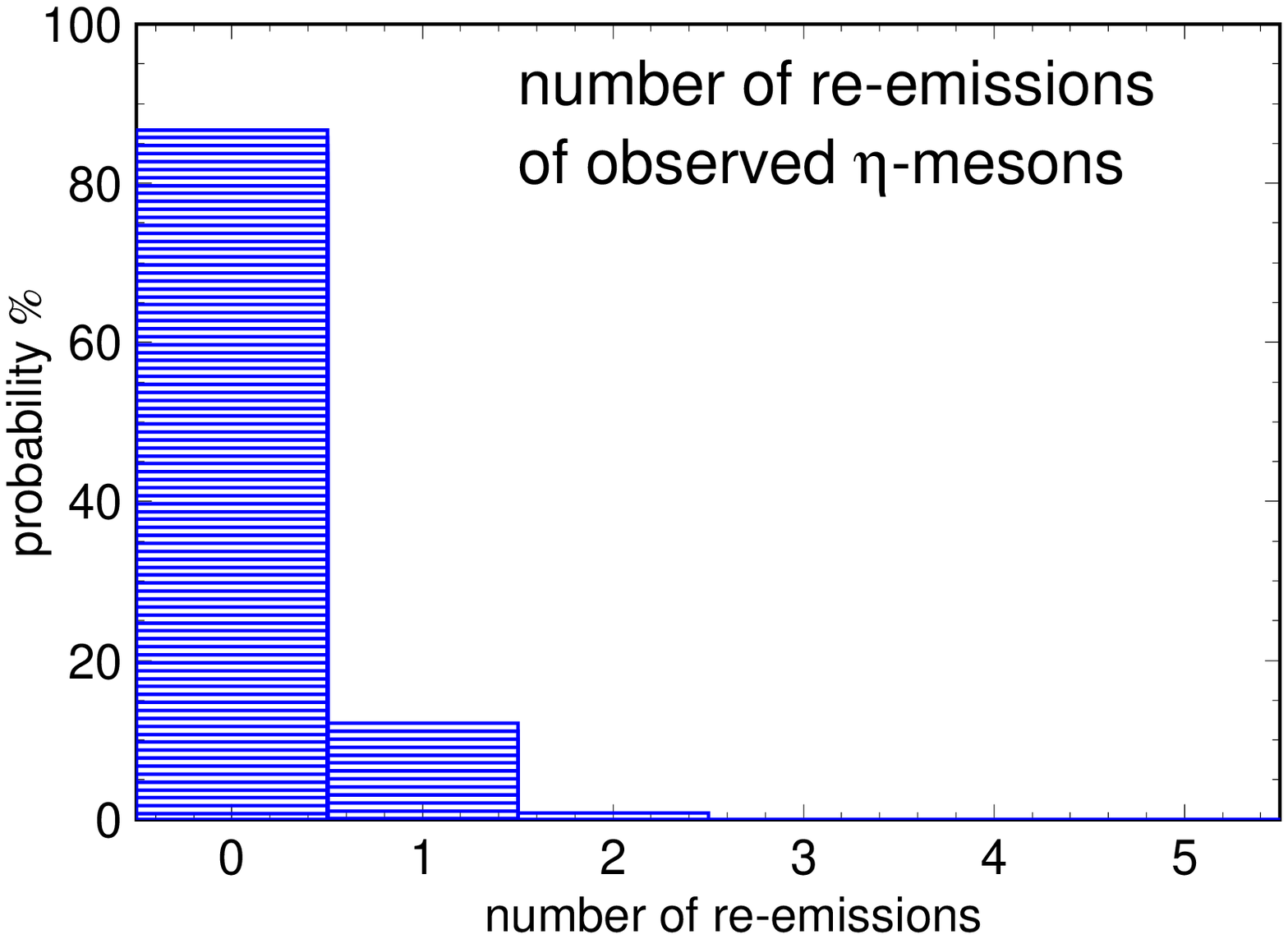}
\end{minipage}
\begin{minipage}{5.2cm}
\caption{Original creation points (left hand side, histograms) and number of 
re-absorptions (right hand side) for pions (upper part) and $\eta$-mesons 
which finally leave a lead nucleus \cite{Hombach_95}. Curves on the right 
hand side correspond to the distribution of nucleons and the nuclear density.
}
\label{fig:16}
\end{minipage}
\end{figure}

\vspace*{0.5cm}
The influence of FSI on pion and $\eta$ production from nuclei is schematically
illustrated in fig. \ref{fig:16} with the help of the BUU calculations
\cite{Hombach_95} for lead. Shown is the distribution of the original creation 
points of observed pions and $\eta$ mesons. Most mesons are emitted from the 
nuclear surface region at densities $\rho\approx\rho_o /2$, where $\rho_o$ is 
the normal nuclear matter density. Pions and $\eta$ mesons are quite
similar in this respect. However, they behave very differently as far as the
number of re-absorption processes for observed mesons is involved. Pions
usually have a long history of propagation through $\Delta$ resonance
formation, while the observed $\eta$ mesons are practically undisturbed.
The reason is, that re-absorbed $\eta$ mesons are almost always lost since the
S$_{11}$ resonance has a 50\% decay branching ratio into $N\pi$. 

\begin{figure}[t]
\centerline{\epsfysize=6.5cm \epsffile{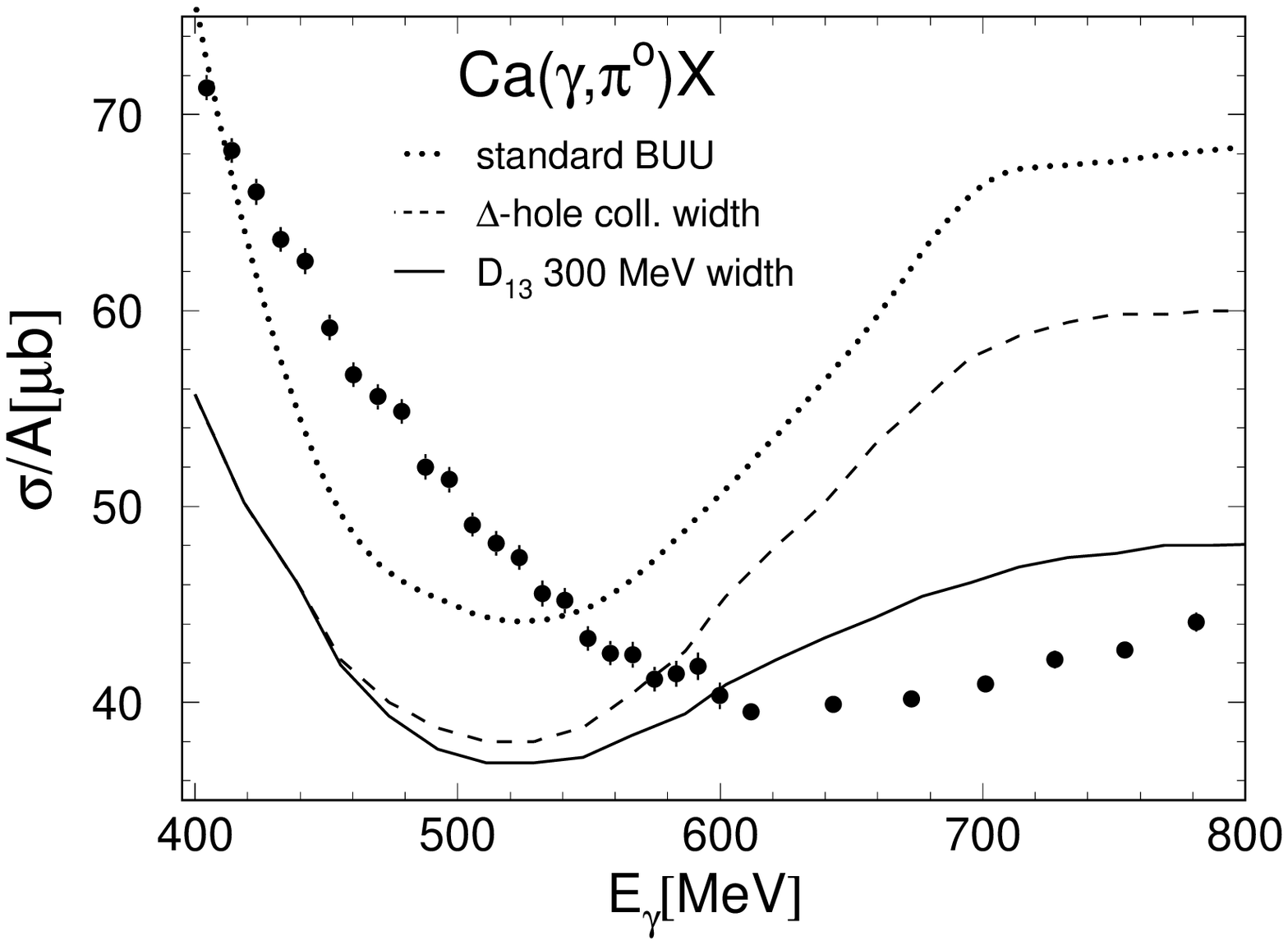}
\hspace*{0.6cm}\epsfysize=6.8cm \epsffile{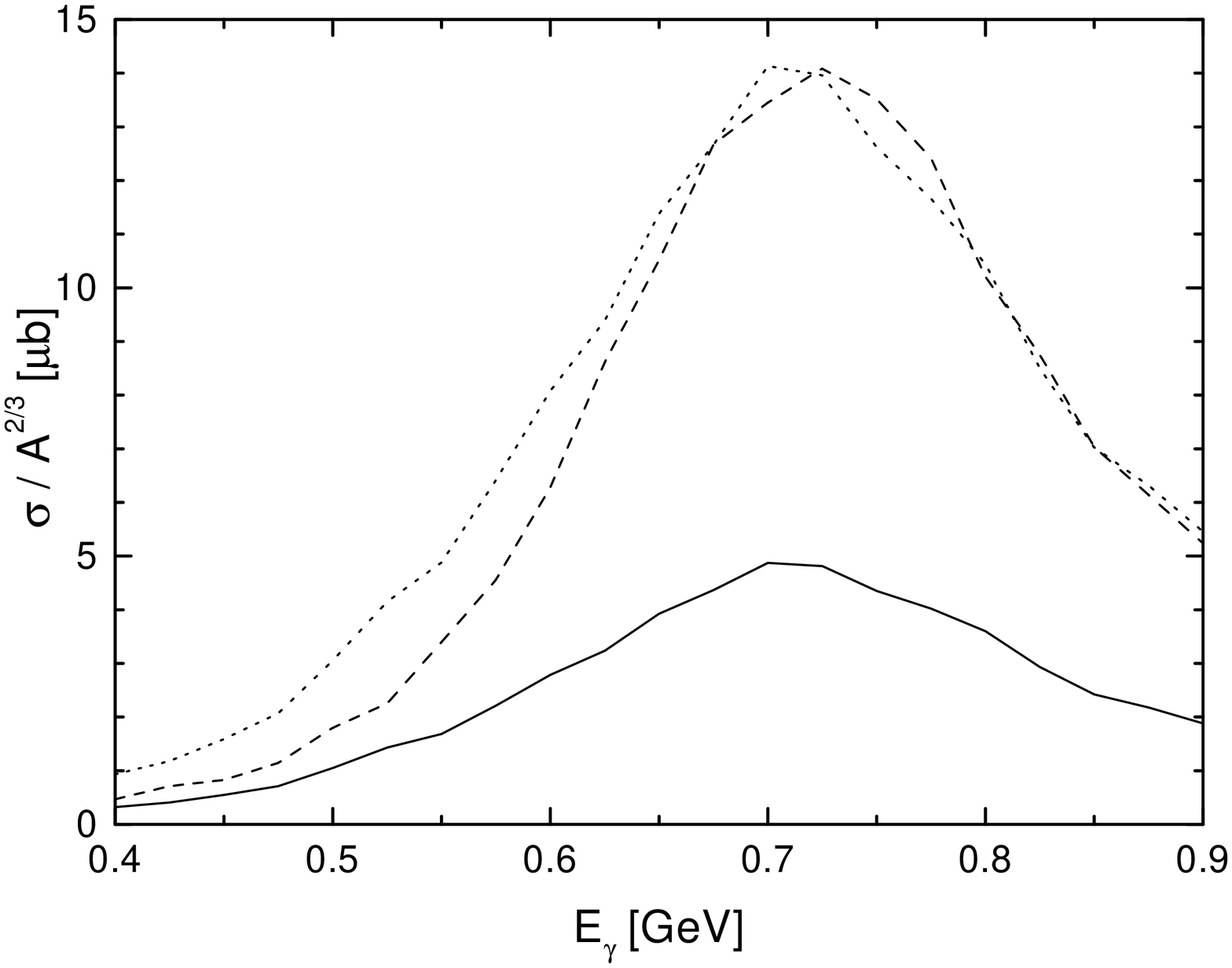}}
\caption{Left hand side: total inclusive $\pi^o$ cross section compared to
BUU model calculations \cite{Krusche_01,Lehr_00}. Right hand side: prediction
for the shape of the D$_{13}$ observed in single $\pi^o$ photoproduction. 
Dashed curve: no in-medium broadening, full curve: 300 MeV
in-medium broadening, dotted curve: full curve scaled up by factor 2.9
\cite{Lehr_01}.  
}
\label{fig:17}
\end{figure}

Although FSI has a large influence on the observed pion spectra, it can not
alone explain the disappearance of the second resonance bump. This can be
demonstrated with a comparison of the data to BUU calculations. The left hand
side of fig. \ref{fig:17} shows the excitation function for inclusive 
$\pi^o$ photoproduction from $^{40}$Ca. BUU model calculations which only
include all the FSI and the in-medium properties of the $\Delta$ produce a much
larger bump in the second resonance region as is observed in experiment.
Only a substantial broadening of the D$_{13}$ resonance brings the model
results closer to the data. The immediate question is, how this result can be
re-concealed with the observation that the width of the measured structure for
the D$_{13}$ resonance, as shown in fig. \ref{fig:13} for single $\pi^o$ 
photoproduction, is basically identical for heavy nuclei and the (quasi)free
nucleon. Lehr and Mosel \cite{Lehr_01} have argued, that this could be due to
a `sampling' effect, which has nothing to do with FSI. The problem is the
following: assume that the resonance is broadened due to modified or additional
decay channels like e.g. $NN^{\star}\rightarrow NN$ (collisional broadening) in 
a density dependent way $\Gamma_{coll}=\Gamma_{coll}^o\times \rho /\rho_o$.
In that case the branching ratio $b_1$ of the resonance into any other decay 
channel with a density independent partial width $\Gamma_1$ becomes density 
dependent since the total width is density dependent. In the simple case with 
only one open decay channel and the additional in-medium collisional width we 
have:
\begin{equation}
b_1 = \frac{\Gamma_1}{\Gamma_1+\Gamma_{coll}^o\times\rho/\rho_o}
\end{equation}
This means that the branching ratio decreases with increasing density. Since 
an experiment always integrates over the density distribution, the exclusive 
reaction channel is dominated by the low density region with the unmodified
resonance. In this way, the resonance does not appear broadened but only 
depleted in strength. This is demonstrated in fig. \ref{fig:17} (right hand 
side) with a BUU calculation of the D$_{13}$ line shape in single $\pi^o$ 
production \cite{Lehr_01}. 

The effect discussed above will of course not occur for a decay
channel which is responsible for the broadening, since then the branching
ratio will rise as function of the density.
This makes it very interesting to study double pion production in the second
resonance region. The analysis of double pion production from the 
free nucleon has shown that a significant contribution to the decay strength
of the D$_{13}$ resonance \cite{Zabrodin_99,Langgaertner_01} comes from the 
D$_{13}\rightarrow N\rho$ decay. The large broadening of the D$_{13}$
in-matter spectral function predicted in \cite{Post_04} is related to this
channel. In double pion production
the $\rho$ contributes to the $\pi^o\pi^{\pm}$ and $\pi^+\pi^-$ final states, 
but not to $\pi^o\pi^o$ since $\rho^o\rightarrow\pi^o\pi^o$ is forbidden.  
This means, that a possible broadening in the observable excitation functions
would be suppressed in $\pi^o\pi^o$ with respect to $\pi^o\pi^{\pm}$.
The measured excitation functions for $\pi^o\pi^o$ and $\pi^o\pi^{\pm}$ from
heavy nuclei are compared to the respective nucleon cross sections in fig.
\ref{fig:18}. 
\begin{figure}[h]
\centerline{\epsfysize=9.3cm \epsffile{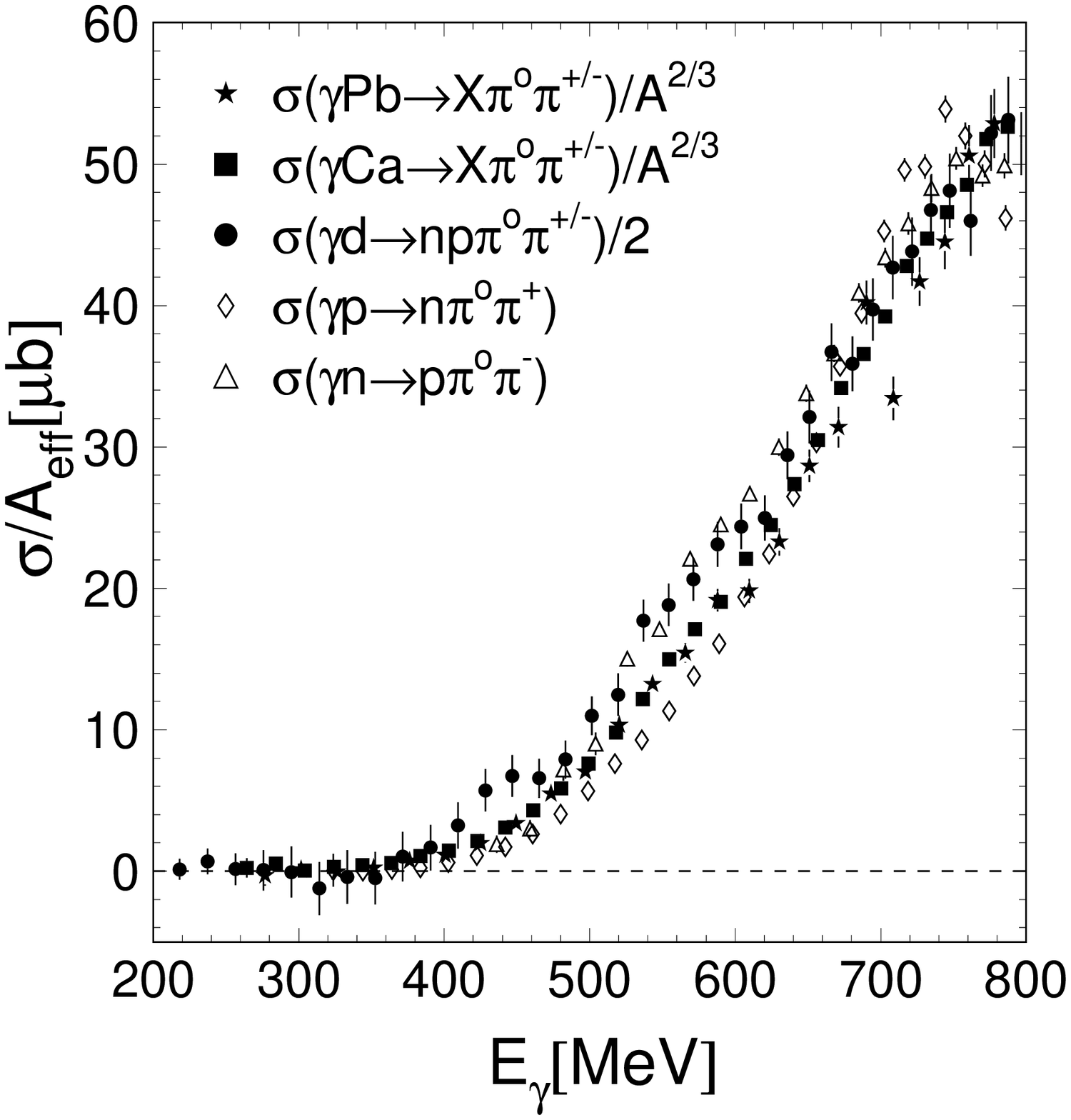}
\epsfysize=9.3cm \epsffile{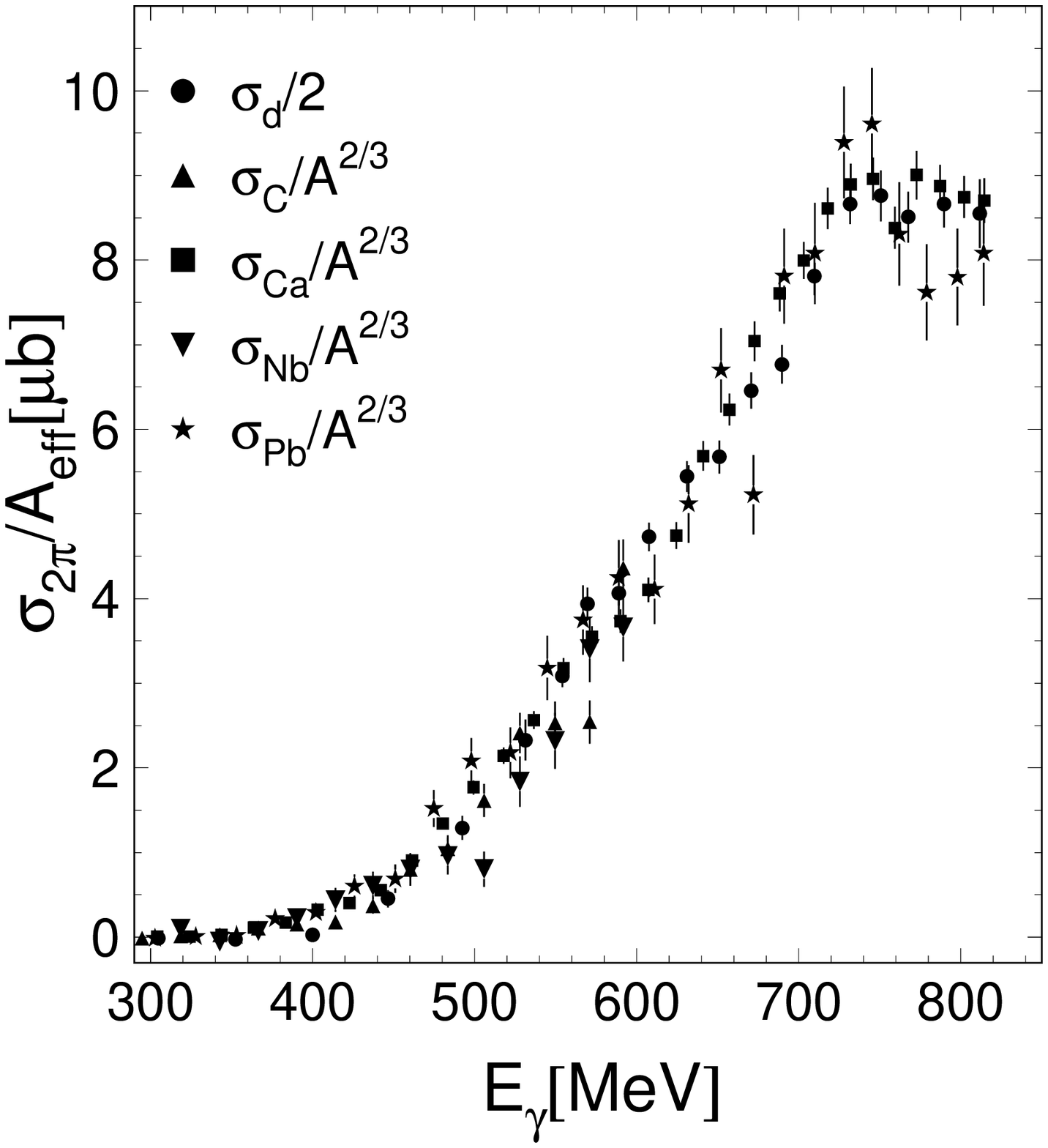}}
\caption{Double pion production from the (quasi)free nucleon and from heavy
nuclei. Left hand side: final states $\pi^o\pi^{\pm}$
\cite{Zabrodin_97,Langgaertner_01,Krusche_04},
right hand side: final state $\pi^o\pi^o$ \cite{Kleber_00,Krusche_04}
}
\label{fig:18}
\end{figure}
The behavior is identical in both cases, the nuclear cross
sections scale with respect to the deuteron following the empirical scaling law:
\begin{equation}
\frac{\sigma_x^{qf}(A)}{A^{2/3}}\approx\frac{\sigma_x^{qf}(d)}{2}
\end{equation}
which holds also for single $\pi^o$ and $\eta$ photoproduction in this energy
region \cite{Krusche_04}. Consequently, there is no evidence for the predicted
broadening of the D$_{13}$ due to the coupling to the $\rho$ meson. The scaling
law of course indicates strong FSI, so that the observed cross sections again
reflect only the conditions in the low density surface region. 

The same is true for the cross section sum $\sigma_{S}$ of all quasifree 
reaction channels with neutral mesons \cite{Krusche_04a}:
\begin{equation}
\sigma_{S} = 
\sigma_{\pi^o}^{qf}+\sigma_{\eta}^{qf}
+\sigma_{2\pi^o}^{qf}+\sigma_{\pi^o\pi^{\pm}}^{qf}
\end{equation}
which is shown in fig. \ref{fig:19} (middle part).
Contributions from coherent single $\pi^o$ production are included into
$\sigma_{\pi^o}^{qf}$ \cite{Krusche_04} and thus also into $\sigma_{S}$.
The behavior of $\sigma_S$ throughout the second resonance is very similar for
the deuteron and the heavy nuclei. The resonance structure is almost 
identical, no in-medium effects are visible, and the scaling indicates the 
dominance of FSI effects.

\begin{figure}[t]
\epsfysize=7.cm \epsffile{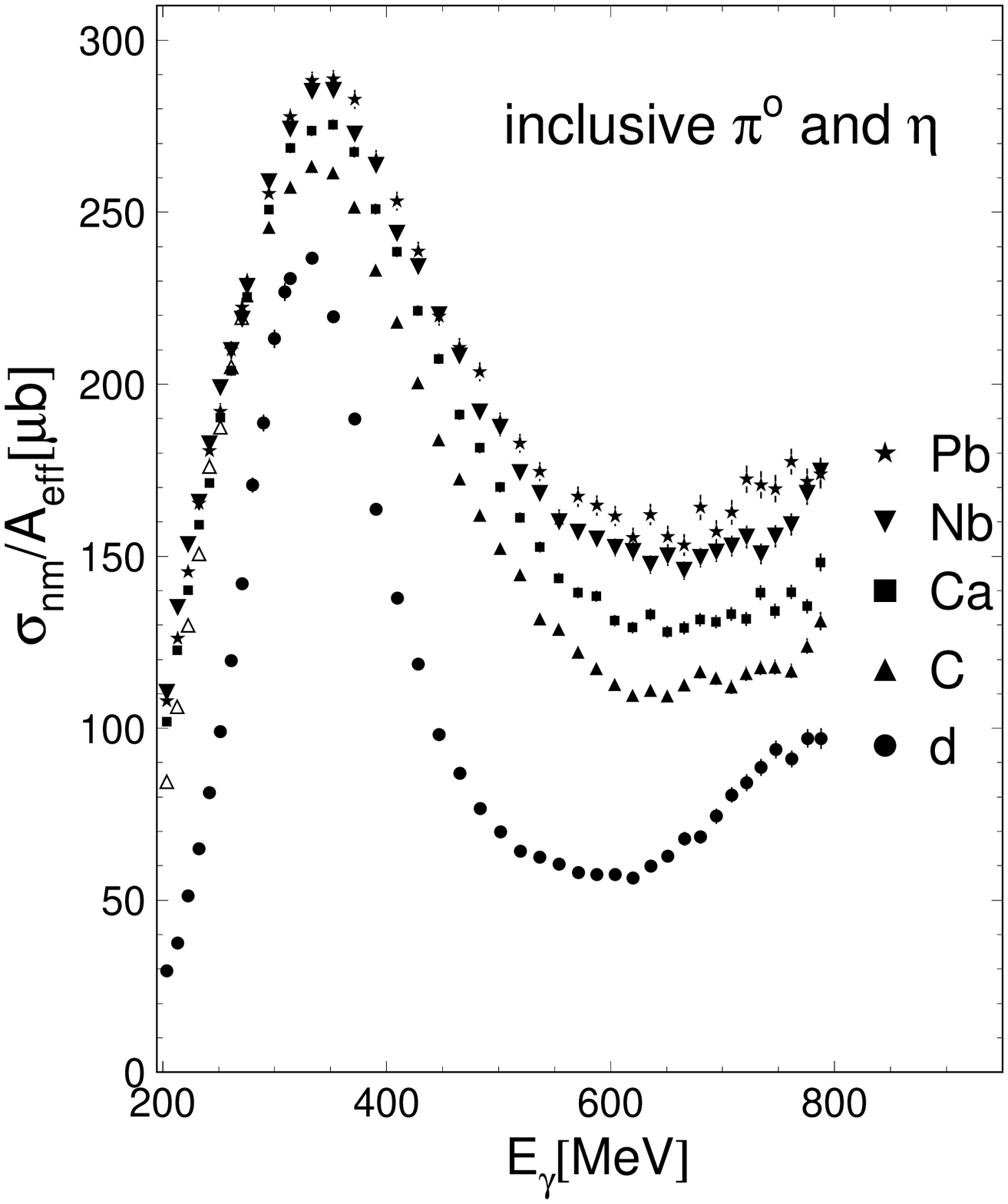}
\epsfysize=7.cm \epsffile{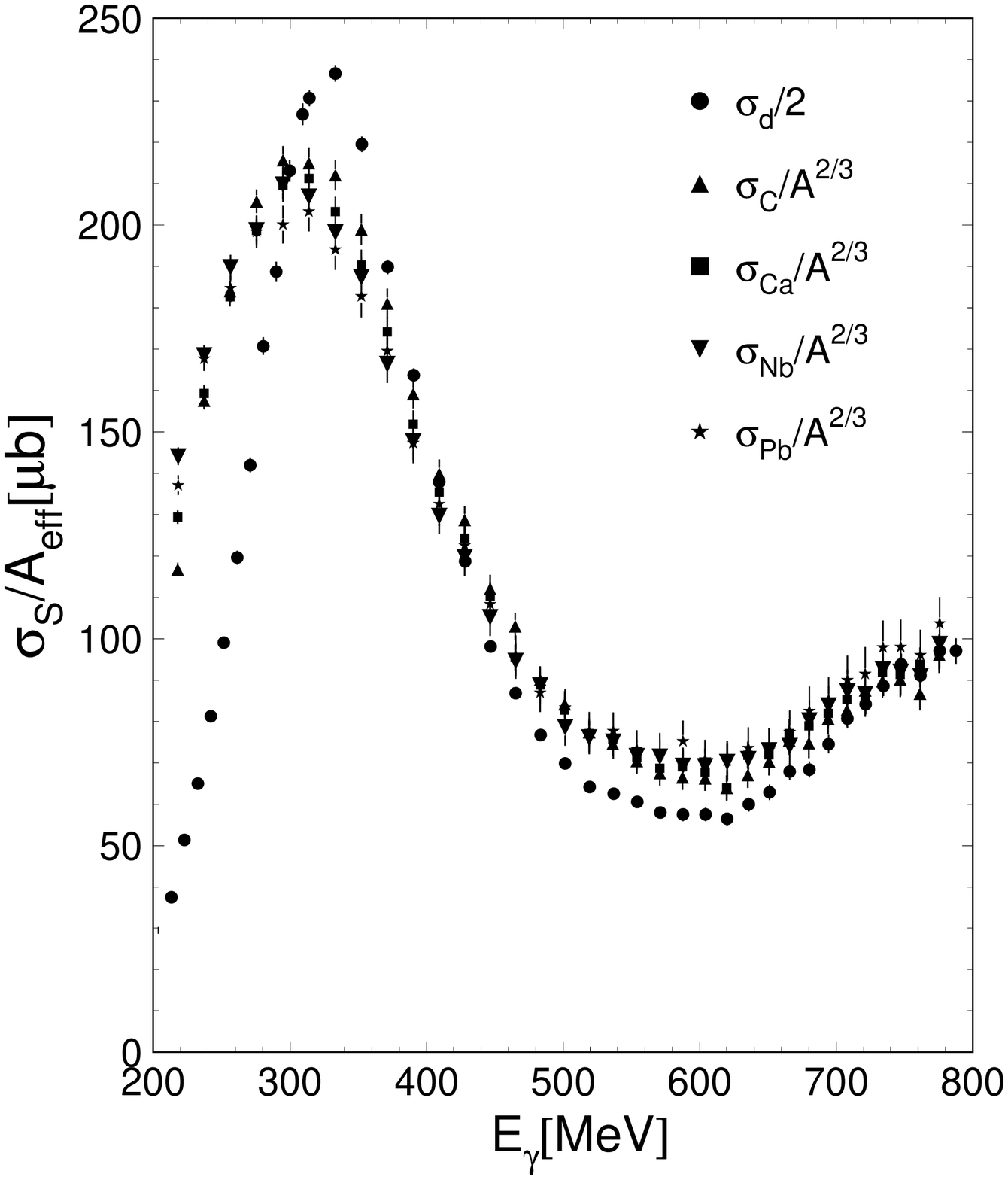}
\epsfysize=7.cm \epsffile{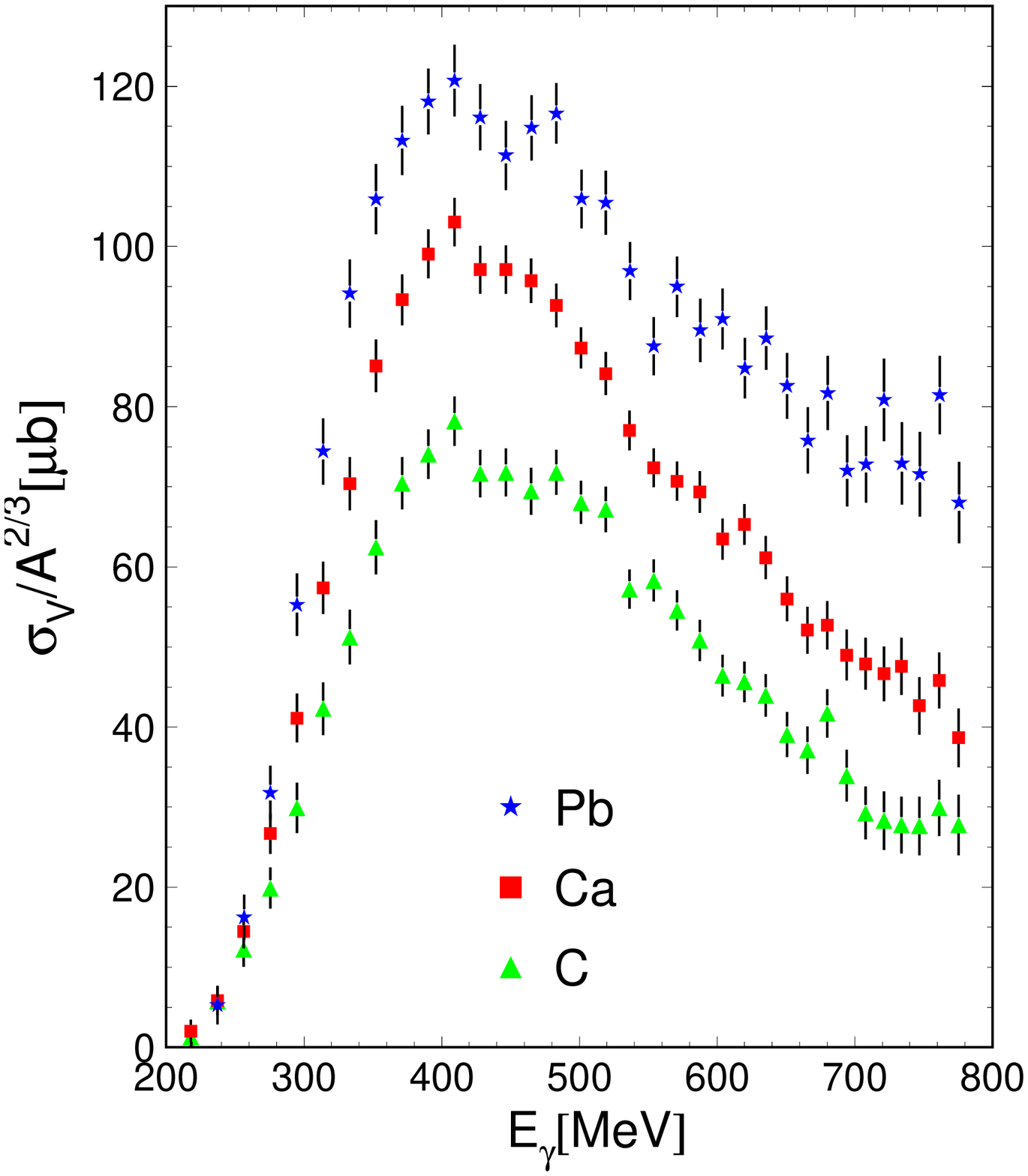}
\caption{Left hand side: total inclusive cross section $\sigma_{nm}$ for 
neutral meson production ($A_{eff}$=2 for the deuteron and $A_{eff}=A^{2/3}$
for $A>2$), middle: sum of exclusive quasifree and coherent
channels $\sigma_{S}$, right hand side: non-quasifree components $\sigma_V$.
}
\label{fig:19}
\end{figure}

The total inclusive cross section of neutral pion and $\eta$ photoproduction
$\sigma_{nm}$ (see fig. \ref{fig:19}, left hand side) was also 
extracted in \cite{Krusche_04}. It includes reactions where for example
a single neutral pion is observed which does not fulfill the kinematic
constraints of quasifree or coherent reactions. These are mainly reactions with
strong FSI, e.g. double pion production with one pion re-absorbed in the
nucleus. The behavior is somewhere in between total photoabsorption and the 
quasifree component. The resonance structure is still visible for heavy nuclei,
but it is much less pronounced than for the deuteron. The difference 
$\sigma_V=\sigma_{nm}-\sigma_S$ between the inclusive cross section 
and the quasifree components (see fig. \ref{fig:19}, right hand side) has
a completely different energy dependence. This part of neutral meson production
from nuclei does not show any indication of the second resonance bump.

The mass number scaling of the different components of the total neutral meson
production cross section follow a simple
\begin{equation}
\label{eq:alpha}
\sigma (A)\propto A^{\alpha}
\end{equation}  
behavior. The results of the fitted exponent $\alpha$ are summarized in fig.
\ref{fig:20} (left hand side). In case of the quasifree component $\sigma_S$
the exponent $\alpha$ is close to 2/3 over the whole energy range. This is 
the expected behavior of surface dominated meson production. However, $\alpha$ 
is significantly larger for the non-quasifree components, in the
second resonance region it approaches even unity, which indicates that this
contribution probes to some extent the nuclear volume. In this case, the 
appearance of the second resonance peak in $\sigma_S$ and its complete
suppression in $\sigma_V$ could indicate a strong density dependence of the
effect. The qualitative behavior of $\alpha$ as function of photon energy 
is reproduced by the BUU calculations \cite{Lehr_00}, however in particular for
$\sigma_S$ the absolute values are underestimated.

A detailed comparison of the different meson production components to the
BUU model results is shown for $^{40}$Ca in fig. \ref{fig:20} (right hand side).
The discrepancy for $\sigma_S$ at low incident photon energies can be
attributed to coherent $\pi^o$ photoproduction which is not included in the
model. The discrepancy at higher incident photon energies is less well
understood, although at least part of it probably comes from two-body
absorption processes of the photon of the type $\gamma NN\rightarrow N\Delta$,
which are also not included in the BUU model. The strength of the second
resonance bump is more or less reproduced for the quasifree part $\sigma_S$,
but is is still significantly overestimated for $\sigma_V$, although the
calculation already includes the strong broadening of the D$_{13}$ resonance. 
In summary, there seems to be evidence that the peak structure in the second
resonance region is unmodified in the low density nuclear surface region but
strongly suppressed in photoproduction reactions which are not entirely 
dominated by the nuclear surface. A quantitative understanding of this effect
is not yet available.  

\begin{figure}[h]
\begin{center}
\epsfysize=8.5cm \epsffile{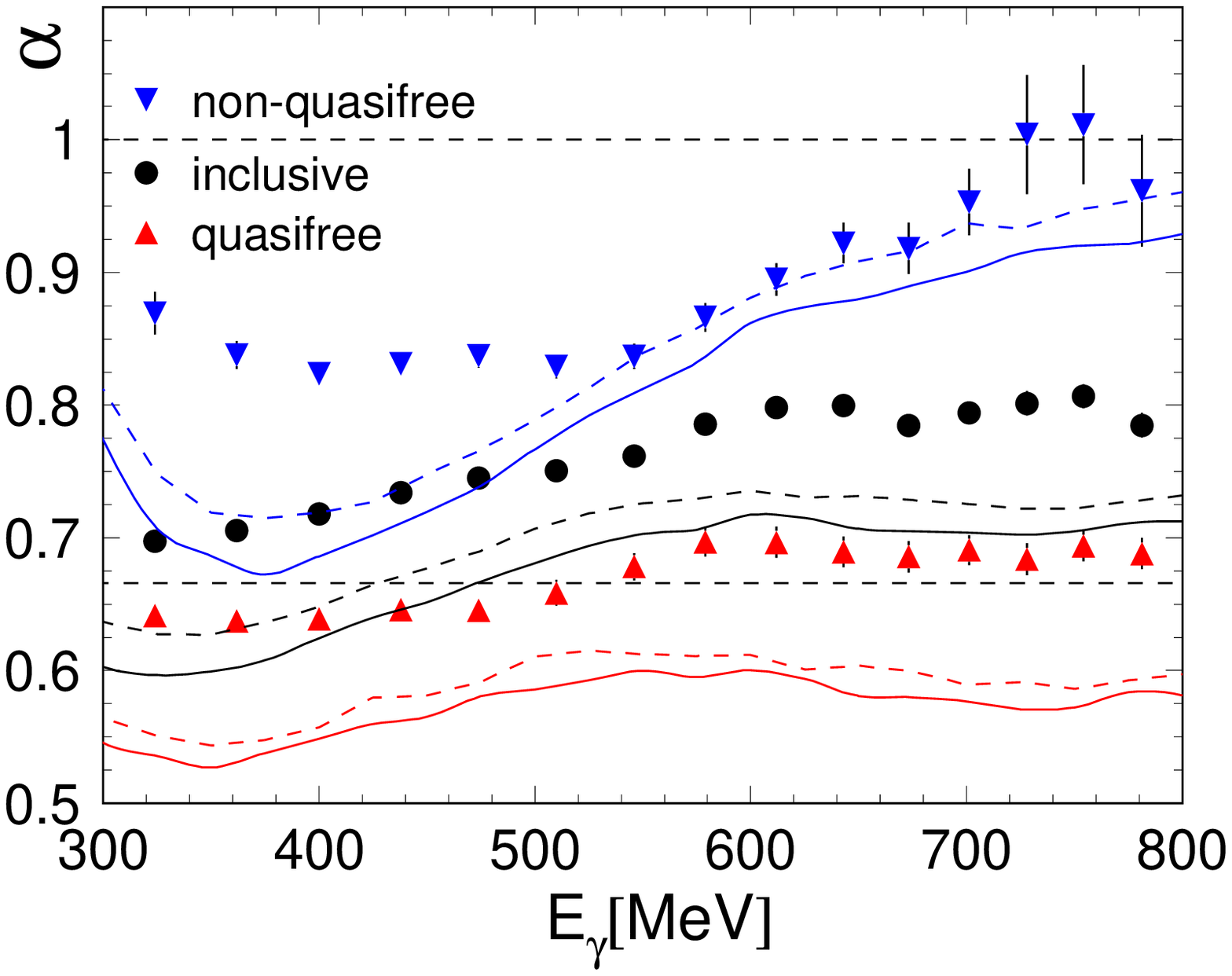}
\epsfysize=8.5cm \epsffile{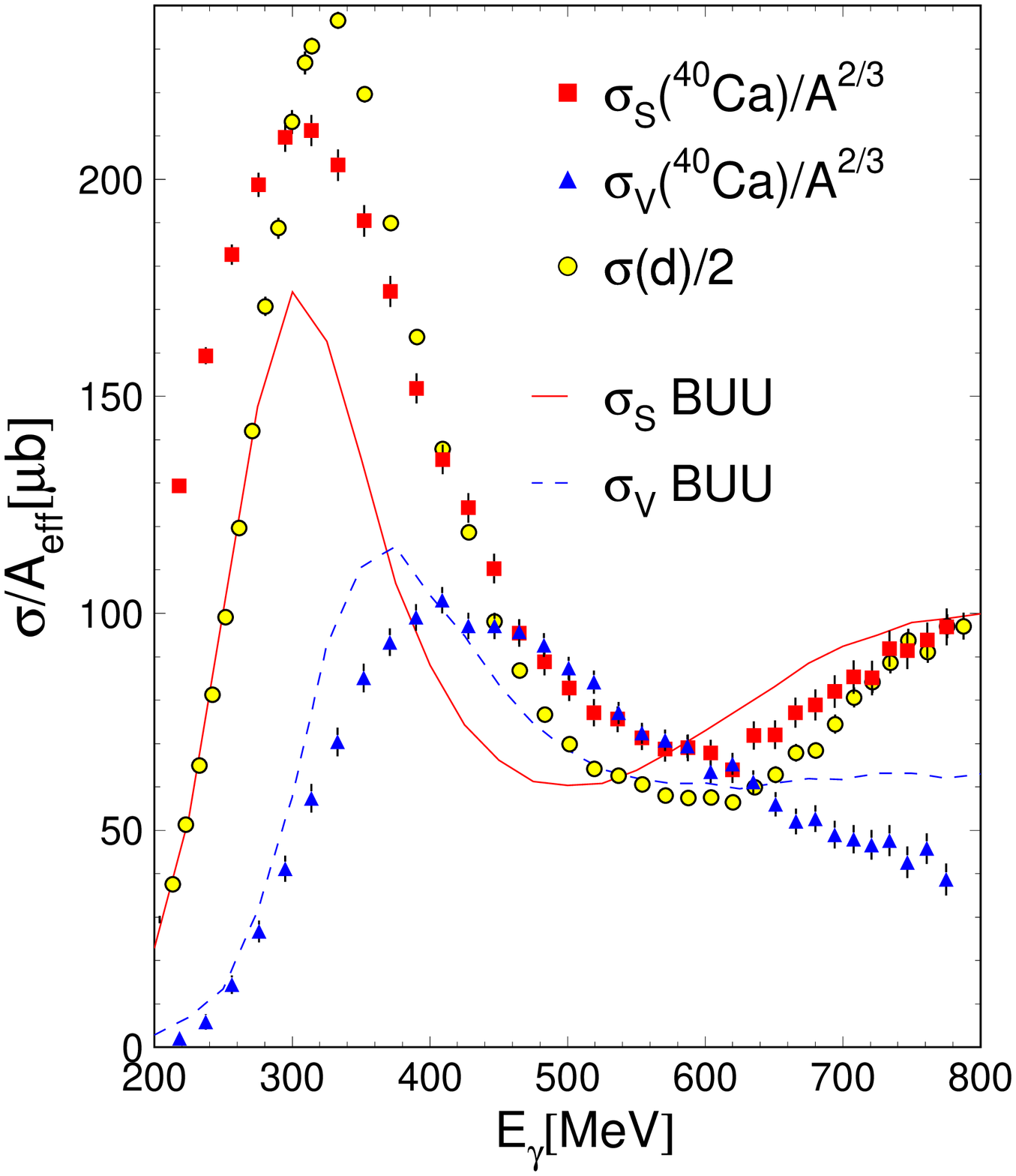}
\caption{Left hand side: 
Scaling of the total cross sections $\sigma_{nm}$, $\sigma_S$ and $\sigma_V$
with mass number as function of incident photon energy. Curves: BUU-model
\cite{Krusche_04,Lehr_00} results with slightly different treatment of the 
P$_{33}$ in-medium width. Right hand side: comparison of $\sigma_S$ and
$\sigma_V$ for $^{40}$Ca to BUU predictions and the cross section for the
deuteron.}
\label{fig:20}
\end{center}
\end{figure}

\subsection{\it $\eta$-mesic nuclei}
The study of the interaction of mesons with nucleons and nuclei has largely
contributed to our understanding of the strong force. In the case of long-lived
mesons like charged pions or kaons, secondary beams can be prepared, which allow
the detailed investigation of such interactions. Much less is known for
short-lived mesons like the $\eta$. Their interaction with nuclei is only
accessible in indirect ways for example when the mesons are first produced
in the nucleus from the interaction of some incident beam and then subsequently
undergo final state interaction (FSI) in the same nucleus. The interaction of
$\eta$-mesons with nuclei is of particular interest because the existence of
bound $\eta$-nucleus systems has been discussed. 
The pion-nucleon interaction at small pion momenta is weak, so that no bound
pion-nucleus states can exit. However, the $\eta$-N interaction at small 
momenta is strongly influenced by the existence of the s-wave nucleon resonance 
S$_{11}$(1535), which lies close to the $\eta$ production threshold and 
couples strongly to the N$\eta$-channel \cite{Krusche_95,Krusche_97}.
An attractive s-wave interaction was already found in coupled channel analysis
of $\eta$ production by Bhalerao and Liu \cite{Bhalerao_85}
($\eta$N scattering length: a=0.27+$i$0.22).
The first suggestion of bound $\eta$-nucleus systems with A $>$ 10
termed {\it $\eta$-mesic} nuclei goes back to Liu and Haider \cite{Liu_86}. 
However, although it was searched in different reactions 
\cite{Chrien_88} for such states, up to now no conclusive evidence 
was reported. Recently, Sokol et al. \cite{Sokol_00} claimed 
the formation of $\eta$-mesic nuclei with mass number $A=11$ (carbon,
beryllium) in the $\gamma +^{12}$C reaction with the decay chain:
\begin{equation}
\gamma + A \rightarrow N_{1} + (A-1)_{\eta} \rightarrow N_{1} +
(N_{2}+\pi)+(A-2)\;\;.
\end{equation}

More recent analyses of the $\eta$N scattering length found larger values of 
its real part which span the entire range from 0.2 - 1.0 and most cluster 
between 0.5 - 0.8 (see e.g. \cite{Ueda_91}). These
results prompted speculations about the existence of light
$\eta$-mesic nuclei, in particular $^2$H, $^3$H, $^3$He, and $^4$He
(see e.g. \cite{Ueda_91}). 
Such states have been sought in experiments investigating the threshold 
behavior of hadron induced $\eta$-production reactions \cite{Calen_96}, 
in particular $pp\rightarrow pp\eta$, $np\rightarrow d\eta$, 
$pd\rightarrow\eta  ^3\mbox{He}$, $\vec{d}d\rightarrow \eta ^4\mbox{He}$, 
and $pd\rightarrow pd\eta$. All reactions show more or less 
pronounced threshold enhancements. However, so far there is no conclusive 
evidence that the final state interaction is strong enough to form quasi-bound 
states. 

\begin{figure}[h]
\begin{minipage}{7.7cm}
\epsfysize=8.5cm \epsffile{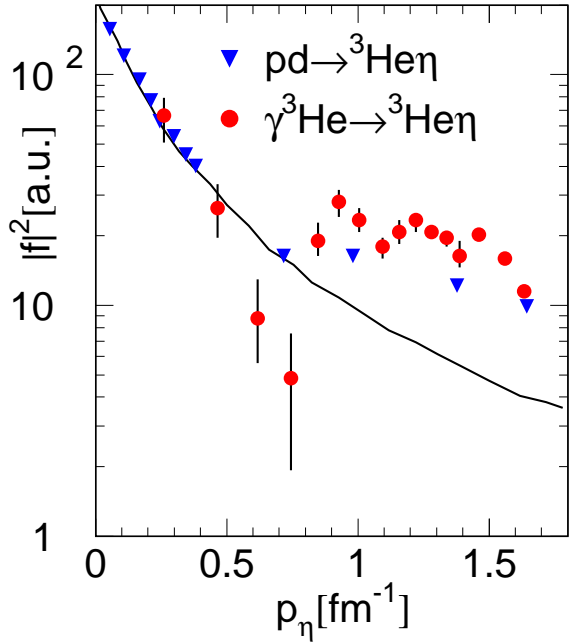}
\caption{Squared amplitudes for proton \cite{Calen_96} and 
photon \cite{Pfeiffer_04} induced $\eta$ production (arb. normalized)
on $^3$He. Solid curve: optical model fit to threshold data
\cite{Calen_96}.}
\label{fig:21}
\end{minipage}
\end{figure}

\vspace*{-11.5cm}
\hspace*{7.9cm}\begin{minipage}{10.cm}
If such states do exist, they should show up as threshold enhancements
independently of the initial state of the reaction. Photoproduction of
$\eta$-mesons from light nuclei was also investigated in detail, in particular
with TAPS at MAMI 
\cite{Krusche_95b,Hoffmann_97,Hejny_99,Weiss_01,Hejny_02,Weiss_03,Pfeiffer_04} 
and again, threshold enhancements were 
observed. These experiments furthermore clearly demonstrated that the reaction 
is dominated by an isovector, spin-flip amplitude (see e.g. \cite{Weiss_03}). 
Consequently, the $I=J$=1/2 nuclei $^3$He and $^3$H are the only light nuclei 
where non-negligible contributions from coherent $\eta$-photoproduction, which
is the ideal channel for the search of near-threshold quasi-bound states, 
can be expected. The coherent reaction was indeed clearly identified for
$^3$He \cite{Pfeiffer_04}, and after reduction of the different phase space
factors, it shows a threshold behavior which is very similar to the 
$pd\rightarrow ^3$He$\eta$ reaction (see fig. \ref{fig:21}).
Since the $^3$He recoil nuclei do not reach the 
detectors, the identification must rely on the different reaction kinematics 
for coherent (final state $\eta +^3$He) and breakup (final states $\eta +pd$ 
or $\eta +ppn$) photoproduction. For this purpose missing energy 
spectra for the $\eta$ mesons were constructed under the assumption of 
coherent kinematics \cite{Pfeiffer_04}. 
\end{minipage}

\vspace*{0.7cm}
In these spectra contributions from coherent production peak around zero 
while contributions from the breakup reactions where the recoil is mainly taken
by one participant nucleon are shifted to negative values. Typical spectra for
the most interesting low energy region are summarized in fig.
\ref{fig:22}.

\begin{figure}[h]
\begin{minipage}{11.4cm}
\epsfysize=7.6cm \epsffile{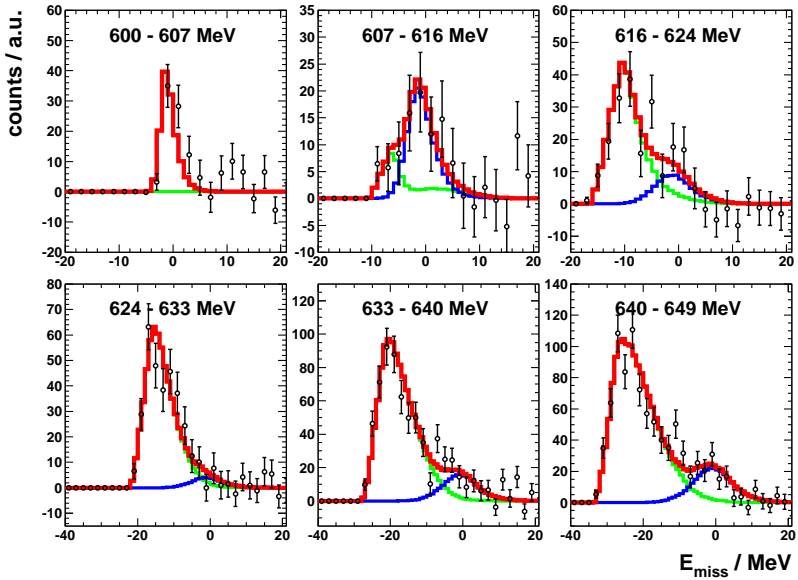}
\end{minipage}
\begin{minipage}{6.8cm}  
\caption{Missing energy spectra assuming coherent reaction kinematics for 
different ranges of incident photon energy. The simulated shapes for the 
coherent (black histograms) and breakup (light grey histograms) parts are 
fitted to the data. The dark grey histograms correspond to the sum of both.
}
\label{fig:22}
\end{minipage}
\end{figure}

\newpage
\begin{figure}[h]
\begin{minipage}{9.5cm}
\epsfysize=7cm \epsffile{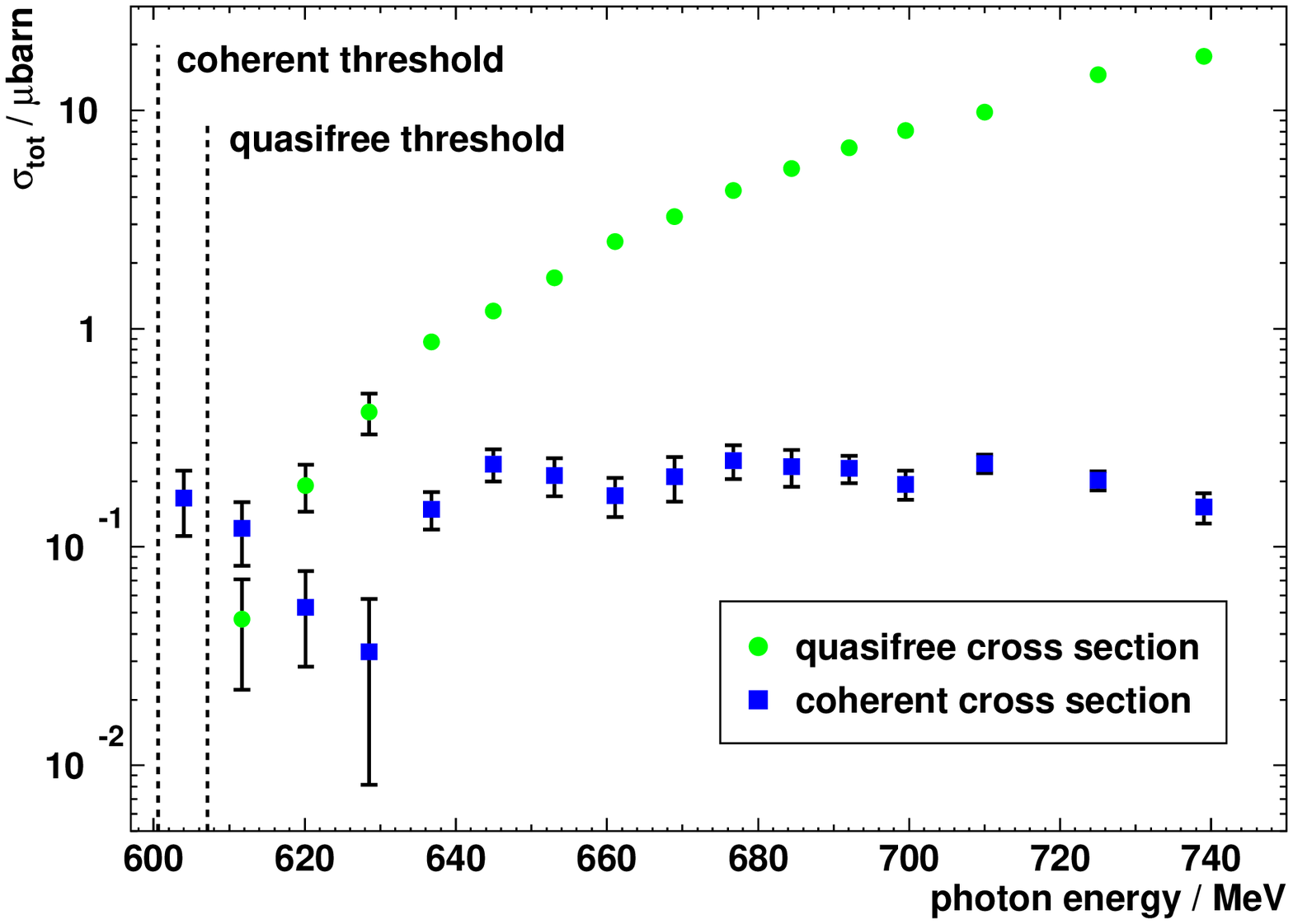}
\epsfysize=3.9cm \epsffile{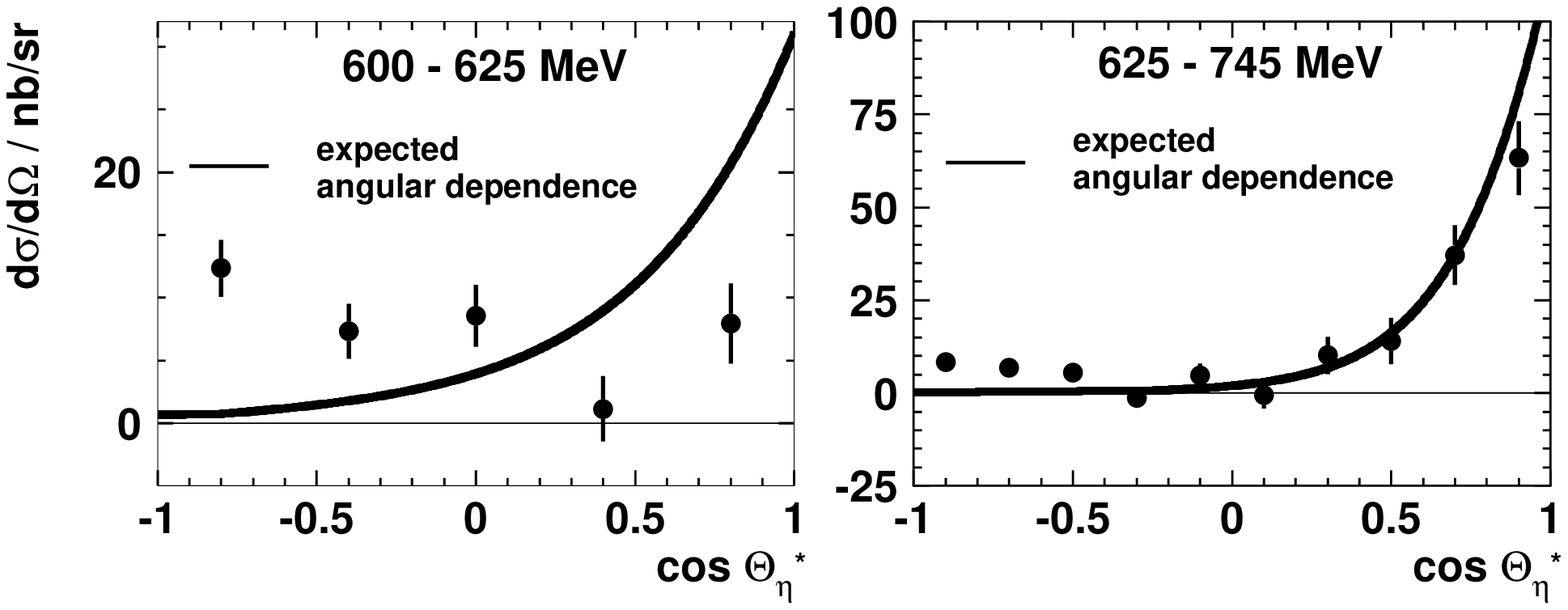}
\caption{Upper part: total cross section for quasi-free and 
coherent $\eta$ production.
Lower part: angular distributions for $^3$He$(\gamma ,\eta)^3$He. 
Solid curves: angular dependence from $^3$He form factor.
}
\label{fig:23}
\end{minipage}
\end{figure}

\vspace*{-13.8cm}
\hspace*{9.5cm}\begin{minipage}{8.3cm}
The threshold behavior of the coherent photon induced reaction 
(see fig. \ref{fig:23}) is remarkable.
In contrary to the breakup process it does not smoothly approach the threshold
but shows a peak-like structure. Furthermore, the angular distribution
in vicinity of the threshold does not follow the behavior expected from the
$^3$He form factor but seems to be much more isotropic. Both could be 
indications for the formation of an intermediate quasi-bound state.  

When an $\eta$ mesic nucleus is formed, the $\eta$ meson can be absorbed on a
nucleon which is excited into the S$_{11}(1535)$ resonance which can 
subsequently decay via pion emission (50\% branching ratio) 
(see fig. \ref{fig:24}). When the state is
populated at incident photon energies below the coherent $\eta$ production
threshold this is basically the only possible decay mode of the system (the
electromagnetic decay of the $\eta$ meson itself is much slower). At energies
above the coherent threshold this channel competes with the emission of $\eta$
mesons. The signature of the additional decay channel are pion - nucleon pairs
which are emitted back-to-back in the rest frame of the $\eta$ mesic nucleus.
\end{minipage}

\vspace*{0.5cm}

\begin{figure}[h]
\begin{minipage}{12.6cm}
\epsfysize=3.2cm \epsffile{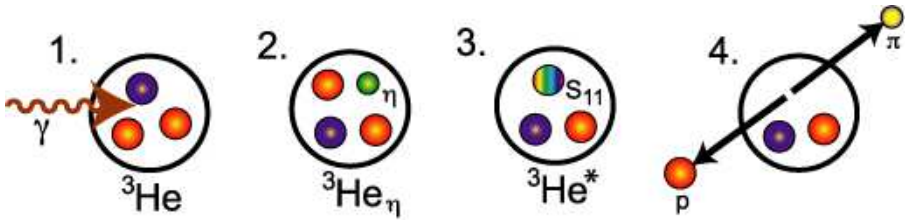}
\end{minipage}
\begin{minipage}{5.5cm}
\caption{Formation of an $\eta$ mesic nucleus and its decay via emission of 
back-to-back nucleon-pion pairs.
}
\label{fig:24}
\end{minipage}
\end{figure}

Such pion - nucleon pairs have been searched for in the channel $\pi^o-p$, 
which is best suited for the TAPS detector. Their excitation function is 
shown in fig. \ref{fig:25}.
Background was estimated by a comparison of the yields for back-to-back 
production (opening angles larger than 170$^o$) to the yield at 
opening angles 150 - 170$^o$. The back-to-back emission shows a structure 
at the production threshold for $\eta$-mesons (600 MeV), which is 
particularly visible in the difference of the two excitation functions 
(see fig. \ref{fig:25}, right hand side). 

It was analyzed \cite{Pfeiffer_04}, if the observed effect in coherent 
$\eta$ production and the structure in the excitation function of pion-nucleon 
emission are roughly consistent with the hypothesis 
that both are different decay channels of an $\eta$ mesic nucleus.
The $\eta$ - mesic (quasi)bound state was parameterized with a Breit-Wigner 
curve at position $W$ with width $\Gamma$.
\begin{figure}[t]
\begin{minipage}{13.5cm}
\epsfysize=4.5cm \epsffile{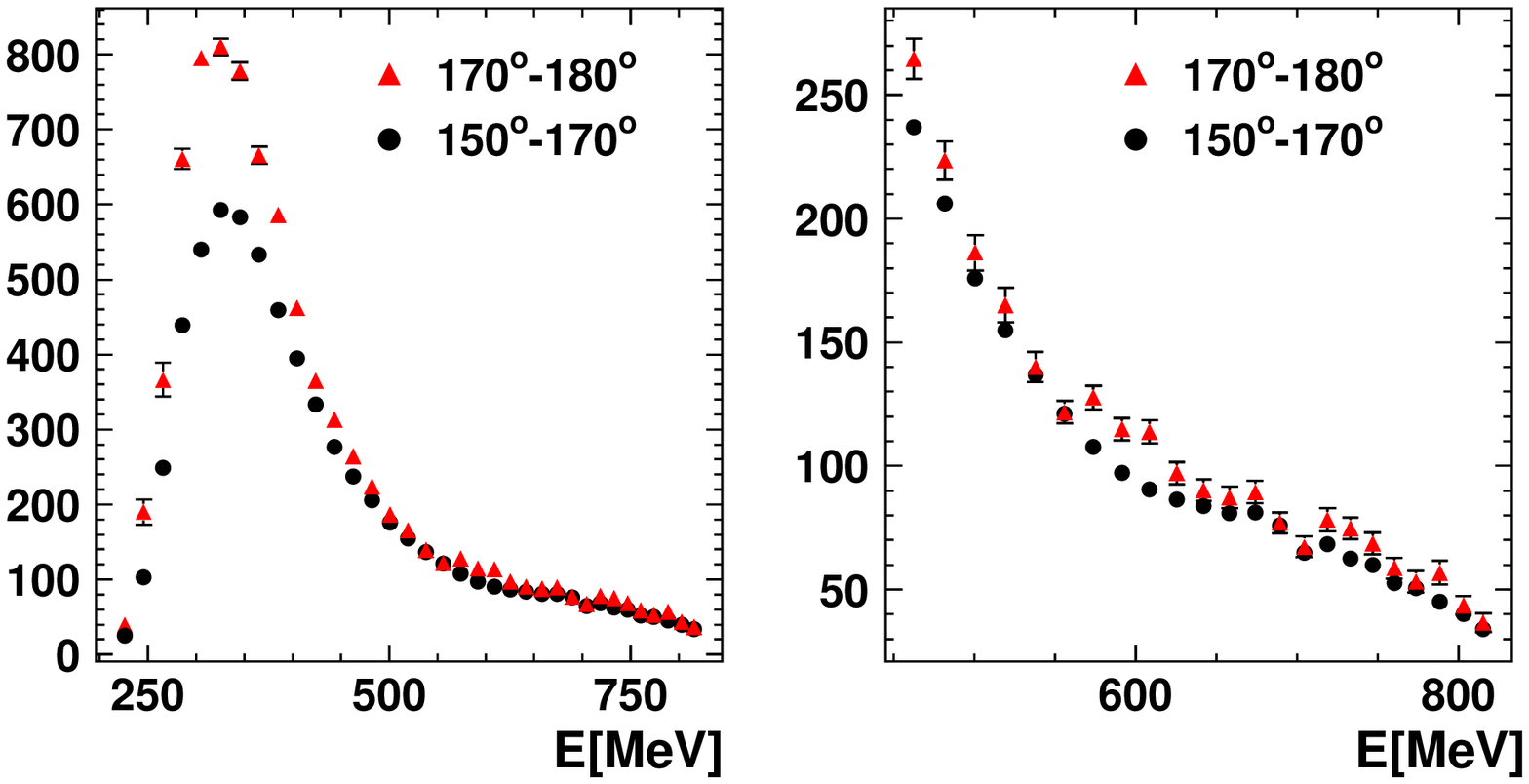}
\epsfysize=4.5cm \epsffile{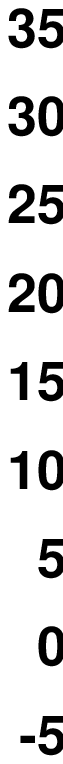}
\end{minipage}
\begin{minipage}{4.8cm}
\caption{Left and center: excitation functions for the $\pi^o-p$ final 
state. Triangles: opening angles between 170 - 180 $^o$, dots: 
150 - 170 $^o$ in the incident photon - $^3$He cm system. 
Right: difference of both distributions fitted with a Breit-Wigner curve.
}
\label{fig:25}
\end{minipage}
\end{figure}
Proper phase space factors and the energy dependent branching ratio of the 
S$_{11}$ resonance have been taken into account \cite{Pfeiffer_04}. The result 
of the fit of this simplified model is shown in fig. \ref{fig:26}. 
A consistent description of the cross sections for both decay channels is 
possible for the following parameters of the Breit-Wigner resonance:
$W = (1481\pm 4) ~\mbox{MeV}$, $\Gamma = (25\pm 6) ~\mbox{MeV}$.
This corresponds to a (quasi)bound state of a width of $\approx$25 MeV,
which is `bound' by $(4\pm 4)$ MeV. 

\begin{figure}[h]
\begin{minipage}{10.cm}
\epsfysize=5cm \epsffile{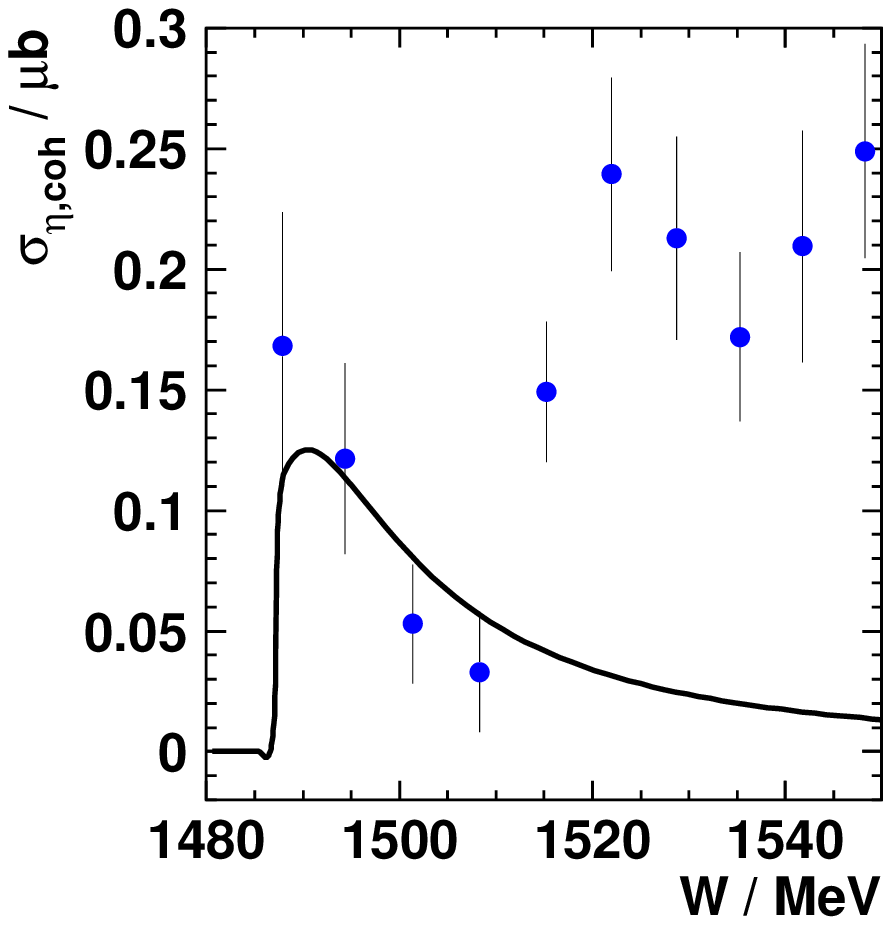}
\epsfysize=5cm \epsffile{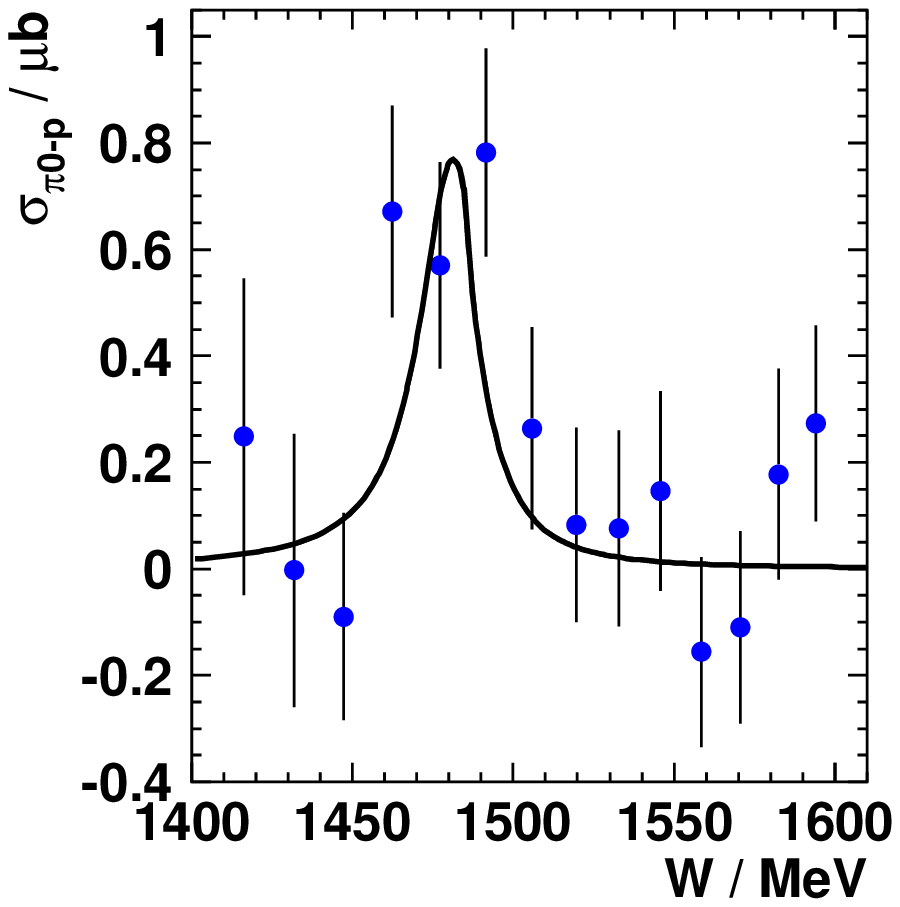}
\caption{Comparison of the two possible decay channels.
Left: coherent $\eta$ channel. Solid curves: best common fit. 
Right: background subtracted excitation function of pion-nucleon back-to-back
pairs. 
}
\label{fig:26}
\end{minipage}
\end{figure}

\vspace*{-7.9cm}
\hspace*{10cm}\begin{minipage}{7.9cm}
However, the statistical significance of 
the signal is still low (3.5 $\sigma$ for the peak in the $\pi^o p$ channel).
Furthermore, Sibirtsev et al. and Hanhart \cite{Sibirtsev_04}
have pointed out that a more detailed analysis of the data does not yet
give solid evidence for a bound state. However, the analysis of Hanhart suggests
that a precise measurement of the width of the peak structure can distinguish
between a bound state and a virtual state. In order to finally solve this
questions a new experiment \cite{Krusche_05} with TAPS and the Crystal Ball 
at MAMI is in preparation. It aims at an improvement of the statistical 
precision by more than an order of magnitude.
\end{minipage}

\vspace*{0.3cm}
\Large
\noindent{{\bf Acknowledgments}}\\
\normalsize
~\\
The discussed results are part of the experimental program of the TAPS
collaboration. I like to acknowledge in particular the contributions of M.
Pfeiffer (eta-mesic nuclei) and F. Bloch, S. Janssen, M. R\"obig-Landau
(heavy nuclei). This work was supported by the Swiss National Fund and the DFG.

\end{document}